\documentclass[preprint,12pt]{elsarticle}

\usepackage{amssymb}
\usepackage{amsmath}
\usepackage{stmaryrd} 
\usepackage{caption} 
\usepackage{subcaption} 
\usepackage{cases} 
\usepackage{bm} 
\usepackage{upgreek} 
\usepackage{url}
\usepackage{multicol} 
\usepackage{xcolor} 

\usepackage{mathtools}
\usepackage{multirow}
\usepackage{algorithm}
\usepackage{algpseudocode}

\begin{document}
\begin{frontmatter}

\title{High-order stabilized matrix-free simulation of rotating mixing devices using the Mortar Element Method}

\author[1]{Bruna Campos}
\ead{bruna.campos@polymtl.ca}

\author[2]{Peter Munch}
\ead{muench@math.tu-berlin.de}

\author[1]{Victor Oliveira Ferreira}
\ead{victor.oliveira-ferreira@polymtl.ca}

\author[1]{Daria Camilla Boffito}
\ead{daria-camilla.boffito@polymtl.ca}

\author[3]{Xavier Banquy}
\ead{xavier.banquy@umontreal.ca}

\author[1]{Bruno Blais\corref{cor2}}
\ead{bruno.blais@polymtl.ca}
\cortext[cor2]{Corresponding author}

\affiliation[1]{organization={Department of Chemical Engineering, Polytechnique Montr\'{e}al},
            addressline={PO Box 6079, Stn Centre-Ville}, 
            city={Montreal},
            postcode={H3C 3A7}, 
            state={Québec},
            country={Canada}}

\affiliation[2]{organization={Institute of Mathematics, Technical University of Berlin},
            addressline={Straße des 17. Juni 135}, 
            city={Berlin},
            postcode={10623},
            country={Germany}}

\affiliation[3]{organization={Faculty of Pharmacy, Universit\'{e} de Montr\'{e}al},
            addressline={PO Box 6128}, 
            city={Montreal},
            postcode={H3C 3J7}, 
            state={Québec},
            country={Canada}}
            
\begin{abstract}
We present a finite element framework to simulate rotating mixing devices using the Mortar Element Method as a domain decomposition strategy. The model is implemented within a matrix-free Navier-Stokes framework which uses a high-order Continuous Galerkin method. The discretized domain is subdivided into rotor and stator parts. An Arbitrary Lagrangian-Eulerian approach accounts for the relative rotor-stator motion, and stabilization is ensured through the Streamline-Upwind/Petrov-Galerkin and Pressure-Stabilizing Petrov-Galerkin methods. The rotor-stator domains are connected by an interface composed of mortar cells, and continuity is weakly enforced in a Discontinuous Galerkin fashion by accounting for boundary integrals at the rotor-stator interface. Verifications of the convergence order in two-dimensional steady and transient examples report optimal rates. The geometric non-conformity created at the mortar interface due to rotor rotation does not introduce significant error in the solution. A three-dimensional example is used to investigate the model’s scalability, which yields ideal strong scaling for large problems. A two-dimensional Rushton impeller example uses a torque analysis to showcase the mesh convergence, and the corresponding velocity profile is in agreement with existing numerical results. In a three-dimensional pitched blade turbine case, the power number curve ($N_p$ vs $Re$) shows good agreement with experimental data for Reynolds number values from 1 to 2000. An energy balance analysis reports a numerical dissipation of 1\% for $Re=200$ and of 10\% for $Re=2000$. By exploiting modern hardware capabilities through matrix-free methods, the proposed model is a robust, accurate, and efficient framework suitable for simulating flows with rotating geometries.
\end{abstract}

\begin{highlights}
\item Novel Continuous-Discontinuous Galerkin model high-order matrix-free framework
\item Geometric non-conformity at the mortar interface does not introduce additional error
\item Verification study shows optimal convergence rates for high-order elements
\item Validation case shows good experimental agreement for Reynolds numbers from 1 to 2000 
\item Energy balance analysis in 3D mixer shows numerical dissipation as low as 1\%
\end{highlights}


\end{frontmatter}


\section{Introduction}
\label{Sec-Intro}

Rotating geometries are a category of engineering systems that encompass various devices such as spinning disc mixers \cite{Oxley2000}, mixing tanks \cite{Blais2016}, and wind turbines \cite{Zahle2009}. Applications of rotating mixing systems comprise, for instance, polymer manufacturing \cite{Rauwendaal1991}, pharmaceutical manufacturing \cite{Zou2021}, and food processing \cite{Cullen2009}. Numerical simulation of this type of device is essential for understanding, monitoring and optimizing the aforementioned processes which are often limited by fluid dynamics or other transport processes which are flow-driven (e.g., mass transfer, heat transfer). Simulations are essential since these systems are difficult to investigate experimentally, considering that they may possess complex geometries and exhibit turbulent flows. From a domain discretization perspective, large mesh deformation can happen due to the rotating motion, and thus remeshing is needed. A feasible alternative to avoid remeshing and maintain accuracy is the Mortar Element Method (MEM), a domain decomposition technique that was initially employed in the context of spectral elements to allow geometric and polynomial non-conformity \cite{Mavriplis1989, Bernardi1993}. It is noteworthy that domain decomposition methods encompass non-overlapping subdomains \cite{Belgacem1997} as well as overlapping meshes for spectral elements  \cite{Merrill2016, Hegazy2026}.

Alternatives to sliding mesh techniques to simulate rotating mixing devices, other approaches comprise (i) rotating frame of reference (RFR) and (ii) the immersed boundary (IB) methods. In the RFR technique, a non-Galilean frame of reference is adopted, and governing equations of fluid motion must account for Coriolis and centrifugal forces \cite{Karimian2025}. Employing multiple reference frames limits the type of flow that can be simulated, since variables must not significantly change in space or time \cite{Joshi2011}. In the IB implementation, the rotating part (e.g., impeller) is discretized separately, and its movement is imposed on the immersed boundary by strategies such as the Nitsche method \cite{Burman2014, Joachim2023} or continuous forcing \cite{Blais2016}. A disadvantage of IB methods is that most of them degrade the order of accuracy of the underlying numerical method. Although some works propose high-order compatible IB \cite{Barbeau2024}, they often yield linear systems with deteriorated condition number. This deterioration can be highly detrimental at higher Reynolds number, where an accurate representation of the boundary layer flow on the blades of the impeller is necessary to obtain a precise prediction of the transient flow structures.

The MEM consists in subdividing the discretized problem into non-over\-lap\-ping subdomains, and then weakly enforcing continuity at the generated interface -- as an alternative to strong point-wise continuity. This interface between subdomains is discretized into the so-called mortar elements \cite{Mavriplis1989}. For rotating mixing devices, two subdomains are generated (rotor and stator), and mortar elements are placed at the rotor-stator interface; the rotor domain contains the rotating element (blades or discs, for instance). Besides employing the MEM for spectral elements \cite{Belgacem1993, Feng2006, Aouadi2014}, other studies extended it to finite element models \cite{Belgacem1997, Abdoulaev1999}, coupling of spectral and finite elements \cite{Belgacem1999, Bernardi1990}, and of finite and boundary elements \cite{Achdou1995}. Mortar methods can be classified according to how the boundary integrals at the subdomains intersection are constructed: using Lagrange multipliers or the Nitsche method. 

The mortar method with Lagrange multipliers has been frequently used in the context of contact problems \cite{Belgacem1998, Puso2004, Wilking2017}, since it provides a variationally consistent form of contact imposition even for non-conforming grids \cite{Farah2018}. Examples include modeling of point, line and surface contact \cite{Farah2018} and large deformation friction \cite{Yang2005}. It was also employed to simulate fluid-structure interactions (FSI) such as slender bodies with fluids \cite{Baaijens2001} and to model one-dimensional fibers into three-dimensional fluid flow \cite{Hagmeyer2024}. Other studies have explored dual-mortar formulations, in which the additional degrees of freedom added by the Lagrange multipliers are condensed \cite{Popp2010, Popp2012}. A drawback from employing Lagrange multipliers is the mesh locking that can become an issue, for instance, when significant difference between refinement of subdomains leads to an over-constrained problem \cite{Sanders2012}.

The Nitsche mortar method (or Nitsche mortaring) uses the interior penalty method to enforce continuity at the mortar interface \cite{Stenberg1998}. Gustafsson \textit{et al.} \cite{Gustafsson2019, Gustafsson2020} employed it for elastic contact problems, comparing formulations with different stabilizing techniques. Bansal \textit{et al.} \cite{Bansal2025} used the Nitsche method for coupling porous media with Navier-Stokes flow within a finite element discretization, and the Nitsche mortaring approach was also used to model incompressible FSI problems in Burman \textit{et al.} \cite{Burman2014}. In the latter study, the Nitsche penalty and the pressure stabilization terms were identified as the main source of consistency error. Formaggia \textit{et al.} \cite{Formaggia2021} modeled moving structures in the FSI context by employing an extended finite element method (FEM) formulation coupled with a Nitsche mortar at the contact interface.

In terms of the finite element approximation, both aforementioned categories of MEM implementation usually rely on a Continuous Galerkin (CG) formulation. Discontinuous Galerkin (DG) formulations, nonetheless, have also been widely explored within the context of, for instance, $hp$-mortar framework \cite{Bertoluzza2016} and sliding meshes for computational fluid dynamics (CFD) \cite{Durrwachter2021}. Kopera \cite{Kopera2015} studied mass conservation for Euler equations and non-conforming meshes for both DG and CG cases. The results between DG and CG formulations were almost identical, and it was shown that non-conformity does not introduce additional errors. Examples of DG models for non-conforming meshes were presented for three-dimensional sliding meshes \cite{Ferrer2012}, compressible Navier-Stokes spectral element model \cite{Durrwachter2021}, and high-order approximation discretization of rotating devices \cite{Laughton2024}.

The relative motion between the rotor and stator domains requires that the underlying numerical model accounts for such motion -- and hence still satisfies its conservation laws. This is commonly done by employing the Arbitrary Lagrangian Eulerian (ALE) framework \cite{Hirt1974, Venkatasubban1995}. It consists of a hybrid method that combines the advantages of both Lagrangian (simple and accurate definition and monitoring of boundaries) and Eulerian (grid movement independent of fluid particle motion) points of view \cite{Duarte2004}. This way, the domain velocity is determined arbitrarily, avoiding constant remeshing -- and possible errors associated with it. The ALE method is often used in FSI problems such as in Dutta \textit{et al.} \cite{Dutta2022}, where a dual-mortar FEM model is proposed, and in Zhao \textit{et al.} \cite{Zhao2025} where a coupled system of incompressible fluid and porous media flow is presented.

Although the MEM is a variationally consistent and highly accurate tool, it requires increased numerical effort due to the necessity of (i) continuous creation of mortar cells via intersection of the stator and rotor meshes, and (ii) computing additional boundary integrals \cite{Mayr2023}. In this sense, efficient parallelization is essential to improve the method's computational cost, especially when simulating large problems. A Message Passing Interface (MPI) parallel implementation for non-conforming grids was discussed by \cite{Krais2021}, and Mayr \& Popp \cite{Mayr2023} implemented a bins system to improve parallel processing. Another strategy used for high-performance computing (HPC) is the use of matrix-free algorithms \cite{Orszag1980}, which take advantage of computer floating-point operations to reduce memory bandwidth limitations of matrix-based solvers \cite{Kronbichler2018}. Recent related works presented high-order matrix-free solvers for the Navier-Stokes in a CG \cite{Saavedra2025} and a DG \cite{Fehn2019} framework. It is noteworthy that high-order FE approximations are useful techniques to reduce the need for excessive mesh refinement by improving accuracy of coarser grids \cite{Wang2013}. Previous works employed high-order methods to simulate sliding domains \cite{Zhang2015, Laughton2024}, DG compressible Navier-Stokes equations \cite{Fehn2019, Durrwachter2021}, and DG spectral formulation for hyperbolic-parabolic systems \cite{Krais2021}. Leveraging MEM implementation with HPC algorithms such as matrix-free methods is a crucial step to counter-balance the computational cost of mortar operators, and this is yet to be presented in the literature.

Given the need of robust, efficient, and accurate models to simulate rotating mixing devices, this work presents a novel model which combines the mortar element method (MEM) with a matrix-free FEM solver. The model is implemented in Lethe \cite{Lethe2025}, an open-source software based on the \texttt{deal.II} library \cite{Arndt2021, Arndt2025} that solves the incompressible Navier-Stokes equations using an implicit Large-Eddy Simulation (LES) approach and a Continuous Galerkin framework. The matrix-free solver described in \cite{Saavedra2025} is employed, and numerical stability is ensured by employing the Streamline-Upwind/Petrov-Galerkin and Pressure-Stabilizing Petrov-Galerkin methods. Relative rotor-stator motion is accounted for through an ALE formulation, and continuity at the rotor-stator interface is enforced by adding boundary terms in a Discontinuous Galerkin fashion with a Symmetric Interior Penalty Galerkin (SIPG) method \cite{Larson2013}. 

A global coarsening geometric multigrid (GCMG) preconditioner \cite{Munch2023} is employed, which is compatible with the matrix-free architecture. To this end, the novelty of our model lies in the combination of the MEM with a matrix-free FEM framework, an approach that overcomes possible memory bandwidth bottlenecks created by using high-order FEM approximations. The computational implementation of the model relies on on-the-fly vector-matrix multiplication in batches, hence taking advantage of modern hardware vectorization and current HPC architectures.

This paper is structured as follows. The governing equations and discretized system of the MEM CFD model are presented in Section~\ref{Sec-GovEqs}, and details on the computational implementation are mentioned in Section~\ref{Sec-Software}. Section~\ref{Sec-Examples} contains steady-state and transient verification cases, a three-dimensional scalability study, and a validation example against experimental results for a pitched blade impeller in a baffled stirred tank. Finally, conclusions and further research are discussed.

\section{Governing equations}
\label{Sec-GovEqs}

The incompressible Navier-Stokes in ALE form for a domain $\Omega$ with a boundary $\partial \Omega$ are given by
\begin{subequations}
\label{EqNS}
\begin{align}
    \nabla \cdot \mathbf{u} &= 0, && (0,T] \times \Omega \label{EqNS:mass} \\
    \partial_t \mathbf{u} + (\mathbf{u} \cdot \nabla) \mathbf{u} - (\mathbf{u}_{\text{ALE}} \cdot \nabla) \mathbf{u} + \nabla p^* - \nu \nabla^2 \mathbf{u} - \mathbf{f} &= 0, && (0,T] \times \Omega. \label{EqNS:momentum}
\end{align}
\end{subequations}
\noindent in which $\mathbf{u}$ is the fluid velocity, $\mathbf{u}_{\text{ALE}}$ is the mesh velocity, $p* = p/\rho$ is the kinematic pressure, $\rho$ is the fluid density, $\nu$ is the kinematic viscosity, and $\mathbf{f}$ is a source term.

The space-discretized form of Eq.~\eqref{EqNS} is obtained using the FEM, where the computational domain $\bar{\Omega}$ (triangulation of $\Omega$) is subdivided into $k$ elements such that $\bar{\Omega} = \cup \: \bar{\Omega}_k$ and $\partial \bar{\Omega}$ is the boundary of $\bar{\Omega}$. By multiplying~\eqref{EqNS:mass} and~\eqref{EqNS:momentum} by test functions $q$ and $\mathbf{v}$, respectively, and summing them, we obtain the following weighted-residual form at the element level:
\begin{align}
\label{EqNSweak}
    (q, \nabla \cdot \mathbf{u})_{\bar{\Omega}_k} + (\mathbf{v}, \partial_t \mathbf{u})_{\bar{\Omega}_k} + (\mathbf{v}, \mathbf{u} \cdot \nabla \mathbf{u})_{\bar{\Omega}_k}- (\mathbf{v}, \mathbf{u}_{\text{ALE}} \cdot \nabla \mathbf{u})_{\bar{\Omega}_k} & \\ \nonumber
    + (\mathbf{v}, \nabla p^*)_{\bar{\Omega}_k} - (\mathbf{v}, \nu \nabla^2 \mathbf{u})_{\bar{\Omega}_k} - (\mathbf{v}, \mathbf{f})_{\bar{\Omega}_k} = 0 &; 
\end{align}
\noindent here we use the following notation for the integrals:
\begin{equation}
    (a, b)_{\bar{\Omega}_k} = \sum_k \int ab \, \text{d}\bar{\Omega}_k.
\end{equation}

\subsection{Mortar element method}
\label{Sec-sub-MortarTerms}

\begin{figure}[t]
\centering
    \includegraphics[scale=0.45]{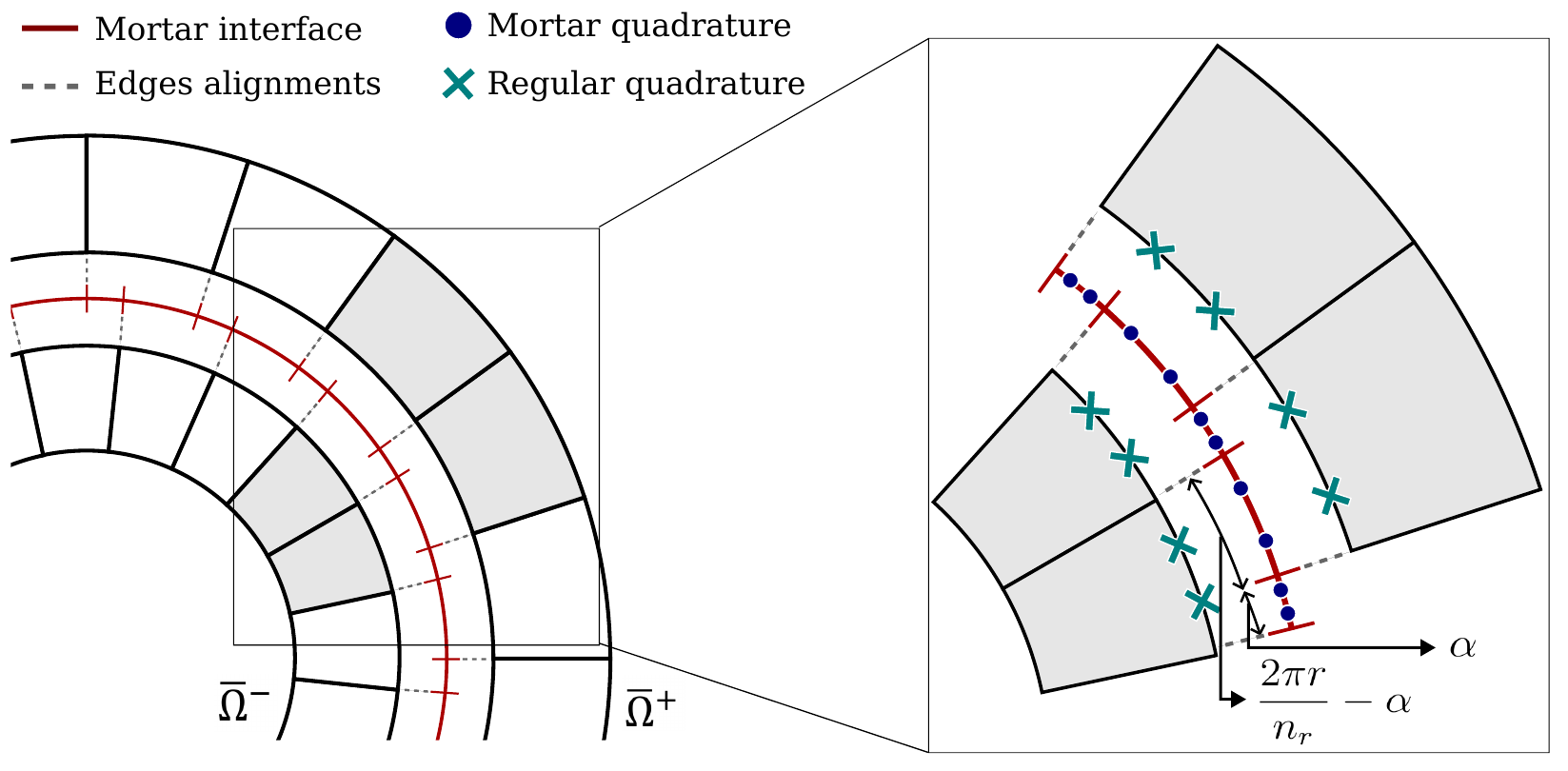}
\caption{Schematic of curved mortar cells (represented in red) at the interface between rotor ($\bar{\Omega}^-$) and stator ($\bar{\Omega}^+$) domains. The mortar interface is partitioned at every intersection with a cell edge, considering both rotor and stator sides. A regular quadrature defined at the boundary cells is mapped into each mortar segment.}
\label{Fig-mortar}
\end{figure}

The subdivision of rotor and stator regions relies on the Mortar Element Method (MEM) \cite{Mavriplis1989, Bernardi1993, Belgacem1999}, which assumes that the discretized domain $\bar{\Omega}$ is subdivided into $n$ non-overlapping subdomains such that
\begin{equation}
    \bar{\Omega} = \cup{\: \bar{\Omega}}^n, \quad \bar{\Omega}^l \: \cap \: \bar{\Omega}^m = \varnothing, \quad \forall \: l, m, \: l \neq m, \quad n = 1, \cdots, N,
\end{equation}
\noindent where $N$ is the total number of subdomains; for the rotor-stator configuration, $N = 2$. In this case, the rotor and stator subdomains are indicated by $\bar{\Omega}^-$ and $\bar{\Omega}^+$, respectively. \textbf{Figure~\ref{Fig-mortar}} depicts the cell configuration at the interface between subdomains; we assume that the rotor and stator sides contain the same number of equally spaced elements. Hence, we allow a non-conforming sliding interface in which each stator cell is subdivided by a rotor cell edge at most once. In a $d$-dimensional domain, the ($d-1$)-dimensional mortar cells are created based on the intersections between the edge of the boundary cells and the mortar interface.

Let us indicate the rotor and stator sides at the mortar interface by $\partial \bar{\Omega}^-$ and $\partial \bar{\Omega}^+$, respectively, such that ${\partial \bar{\Omega}}^*_k = \partial \bar{\Omega}_k \cap(\partial \bar{\Omega}^- \cup \partial \bar{\Omega}^+)$ refers to element boundaries coincident with the rotor-stator interface and ${\partial \bar{\Omega}}^*_{k, \text{all}}$ are all the element boundaries of cells at the interface. We define the average $\{\!\!\{ \cdot \}\!\!\}$ and jump $\llbracket \cdot \rrbracket$ operators of a generic field $\xi$ as:
\begin{equation}
    \{\!\!\{ \xi \}\!\!\} = \dfrac{\xi^{-} + \xi^{+}}{2}, \qquad    \llbracket \xi \rrbracket = \xi^{+} - \xi^{-}.
\end{equation}

The boundary integrals are implemented based on the DG method for incompressible flows proposed by Fehn \textit{et al.} \cite{Fehn2018, Fehn2021}. Continuity on the velocity and pressure fields is ensured by integrating by parts the pressure and viscous terms in~\eqref{EqNSweak}; for the former, using a central approximation of the numerical flux, we have:
\begin{align}
    (\mathbf{v}, \nabla p^*)_{\bar{\Omega}_k} \coloneq - (\nabla \mathbf{v}, p^*)_{\bar{\Omega}_k} + (\mathbf{v}, \{\!\!\{ p^* \}\!\!\} \mathbf{n})_{{\partial \bar{\Omega}}^*_k}.
\label{EqMortarPre}
\end{align}

For the viscous term, we employ the symmetric interior penalty Galerkin (SIPG) method \cite{Arnold1982, Arnold2002} to obtain the following weak form:
\begin{align}
\label{EqMortarDiff}
    - (\mathbf{v}, \nu \nabla^2 \mathbf{u})_{\bar{\Omega}_k} & \coloneq (\nabla \mathbf{v}, \nu \nabla \mathbf{u})_{\bar{\Omega}_k} - (\nabla \mathbf{v}, \dfrac{\nu}{2} \llbracket \mathbf{u} \rrbracket \otimes \mathbf{n})_{{\partial \bar{\Omega}}^*_k} - (\mathbf{v}, \nu \{\!\!\{ \nabla \mathbf{u} \}\!\!\} \cdot \mathbf{n})_{{\partial \bar{\Omega}}^*_k} \nonumber \\ 
    & + (\mathbf{v}, \nu \tau_k \llbracket \mathbf{u} \rrbracket)_{{\partial \bar{\Omega}}^*_k},
\end{align}
\noindent where the penalty parameter $\tau_k$ for each element is defined according to the work of Hillewaert \cite{Hillewaert2013}:
\begin{equation}
    \tau_k = F(\mathcal{P}+1)^2 \dfrac{A({\partial \bar{\Omega}}^*_{k, \text{all}} \backslash {\partial \bar{\Omega}}^*_k)/2 + A({\partial \bar{\Omega}}^*_k)}{V(\bar{\Omega}_k)}.
\label{EqPenalty}
\end{equation}
\noindent In this expression, $F$ is a factor that controls the amount of penalization (generally chosen in the range of 1 to 10 \cite{Larson2013}), $\mathcal{P}$ is the polynomial degree of the FE approximation, $V(\bar{\Omega}_k) = \int \text{d} \bar{\Omega}_k$ is the element volume, and $A$ is the surface area of the interior (${\partial \bar{\Omega}}^*_{k, \text{all}} \backslash {\partial \bar{\Omega}}^*_k$) and exterior (${\partial \bar{\Omega}}^*_k$) element boundaries.

Regarding the mass conservation term, we integrate it by parts twice and adopt the strong DG formulation to improve mass conservation given by \cite{Fehn2018}:
\begin{equation}
    (q, \nabla \cdot \mathbf{u})_{\bar{\Omega}_k} \coloneq (q, \nabla \cdot \mathbf{u})_{\bar{\Omega}_k} - (q, \dfrac{1}{2} \llbracket \mathbf{u} \rrbracket \cdot \mathbf{n})_{{\partial \bar{\Omega}}^*_k}.
\label{EqMortarMass}
\end{equation}

\noindent Substituting relations~\eqref{EqMortarPre}, ~\eqref{EqMortarDiff}, and~\eqref{EqMortarMass} into Eq.~\eqref{EqNSweak}, we have:
\begin{align}
\label{Eq-NSweakWmortar}
    F_{NS}^* := & (q, \nabla \cdot \mathbf{u})_{\bar{\Omega}} - (q, \dfrac{1}{2} \llbracket \mathbf{u} \rrbracket \cdot \mathbf{n})_{{\partial \bar{\Omega}}^*} + (\mathbf{v}, \partial_t \mathbf{u})_{\bar{\Omega}} + (\mathbf{v}, \mathbf{u} \cdot \nabla \mathbf{u})_{\bar{\Omega}} \\ \nonumber
    & - (\mathbf{v}, \mathbf{u}_{\text{ALE}} \cdot \nabla \mathbf{u})_{\bar{\Omega}} - (\nabla \mathbf{v}, p^*)_{\bar{\Omega}} + (\mathbf{v}, \{\!\!\{ p^* \}\!\!\} \mathbf{n})_{{\partial \bar{\Omega}}^*} + (\nabla \mathbf{v}, \nu \nabla \mathbf{u})_{\bar{\Omega}} \\ \nonumber
    & - (\nabla \mathbf{v}, \dfrac{\nu}{2} \llbracket \mathbf{u}  \rrbracket \otimes \mathbf{n})_{{\partial \bar{\Omega}}^*} - (\mathbf{v}, \nu \{\!\!\{ \nabla \mathbf{u} \}\!\!\} \cdot \mathbf{n})_{{\partial \bar{\Omega}}^*} + (\mathbf{v}, \nu \tau_k \llbracket \mathbf{u} \rrbracket )_{{\partial \bar{\Omega}}^*} - (\mathbf{v}, \mathbf{f})_{\bar{\Omega}}.
\end{align}

\subsection{Numerical stabilization}
\label{Sec-sub-Stabilization}

We ensure the stabilization of the weak form by using the Streamline-Upwind/Petrov-Galerkin (SUPG) and Pressure-Stabilizing Petrov-Galerkin (PSPG) approaches \cite{Lethe2025}, and the terms $a_{\text{SUPG}}, \: a_{\text{PSPG}}$ are added into the weak form:
\begin{align}
    \hat{F}_{NS}^* := F_{NS}^* + \underbrace{(\tau_{\mathbf{u}} (\mathbf{u} - \mathbf{u}_{\text{ALE}}) \cdot \nabla \mathbf{v}, \mathcal{R}_{\mathbf{u}})_{\bar{\Omega}_k}}_{a_{\text{SUPG}}} + \underbrace{(\tau_{\mathbf{u}} \nabla q, \mathcal{R}_{\mathbf{u}})_{\bar{\Omega_k}}}_{a_{\text{PSPG}}}.
\label{Eq-NSweakComplete}
\end{align}

The strong residual of the momentum conservation $\mathcal{R}_{\mathbf{u}}$ is defined as:
\begin{align}
    \mathcal{R}_{\mathbf{u}} = \partial_t \mathbf{u} + (\mathbf{u} \cdot \nabla) \mathbf{u} - (\mathbf{u}_{\text{ALE}} \cdot \nabla) \mathbf{u}  + \nabla p^* - \nu \nabla^2 \mathbf{u} - \mathbf{f},
\label{Eq-strong-residual}
\end{align}
\noindent and $\tau_{\mathbf{u}}$ is a stabilization parameter defined according to Tezduyar \cite{Tezduyar1991}:
\begin{equation}
    \tau_{\mathbf{u}} = \left[ \left(\dfrac{1}{\Delta t} \right)^2 + \left( \dfrac{2 || \mathbf{u}- \mathbf{u}_{\text{ALE}}||}{h_{\text{conv}}} \right)^2 + 9\left( \dfrac{4 \nu}{h^2_{\text{diff}}} \right)^2 \right]^{-1/2}
\label{Eq-stab-param}
\end{equation}
\noindent where $\Delta t$ is the time step and $h_{\text{conv}} = h_{\text{diff}} = h$ is the element size given by the diameter of the corresponding area-equivalent circle.

\subsection{Finite element discretization}
\label{Sec-sub-FE}

Considering the Galerkin method, the weak form in Eq.~\eqref{Eq-NSweakWmortar} can be discretized in space by employing the following representation of $\mathbf{v}, \:\mathbf{u}, \: q, \: p^*$:
\begin{equation}
    \mathbf{v} = \sum_{i=1}^{N_u} v_i \, \boldsymbol{\phi}_i, \qquad
    \mathbf{u} = \sum_{j=1}^{N_u} u_j \, \boldsymbol{\phi}_j, \qquad
    q = \sum_{k=1}^{N_p} q_k \, \psi_k, \qquad
    p^* = \sum_{l=1}^{N_p} p^*_l \, \psi_l,
\label{Eq-shape-functions}
\end{equation}
\noindent in which $\boldsymbol{\phi}_i, \boldsymbol{\phi}_j, \psi_k, \psi_l$ are Lagrange polynomials of degree $\mathcal{P}$, and $v_i, \: u_j, \: q_k, \: p^*_l$ are nodal values. The nonlinear system is solved using Newton's method, where a linear system of equations of the form:
\begin{equation}
    \boldsymbol{\mathcal{J}} [\delta \mathbf{u}, \delta p^*]^\top = - \boldsymbol{\mathcal{R}}
\label{Eq-Newton-method}
\end{equation}
\noindent is solved. The weak form of the Jacobian and residual are given by $\boldsymbol{\mathcal{J}}$ and $\boldsymbol{\mathcal{R}}$, respectively, and $[\delta \mathbf{u}, \delta p^*]^\top$ are the computed increments. Writing the linear system in matrix form, we have that:
\begin{equation}
    \begin{bmatrix}
    \hat{\mathbf{A}} & \hat{\mathbf{B}} \\
    \hat{\mathbf{C}} & \hat{\mathbf{D}}
    \end{bmatrix} 
    \begin{bmatrix}
    \delta \mathbf{u} \\
    \delta p^*
    \end{bmatrix} = -
    \begin{bmatrix}
    \mathbf{R}_{\mathbf{u}} \\
    \mathbf{R}_{p}
    \end{bmatrix}.
\label{Eq-Newton-matrix}
\end{equation}

The stabilized Jacobian sub-matrices are given by:
\begin{align}
    \hat{A}^k_{ij} &= A^k_{ij} \nonumber \\
    & + (\tau_{\mathbf{u}}(\mathbf{u} - \mathbf{u}_{\text{ALE}}) \cdot \nabla \boldsymbol{\phi}_i, \partial_t \boldsymbol{\phi}_j + \boldsymbol{\phi}_j \cdot \nabla \mathbf{u} + (\mathbf{u} - \mathbf{u}_{\text{ALE}}) \cdot \nabla \boldsymbol{\phi}_j - \nu \nabla^2 \boldsymbol{\phi}_j)_{\bar{\Omega}_k} \nonumber \\
    & + (\tau_{\mathbf{u}}(\boldsymbol{\phi}_j - \mathbf{u}_{\text{ALE}}) \cdot \nabla \boldsymbol{\phi}_i, \partial_t \mathbf{u} + (\mathbf{u} - \mathbf{u}_{\text{ALE}}) \cdot \nabla \mathbf{u} - \nu \nabla^2 \mathbf{u} + \nabla p^* - \mathbf{f})_{\bar{\Omega}_k}, \\
    \hat{B}^k_{ij} &= B^k_{ij} + (\tau_{\mathbf{u}} (\mathbf{u} - \mathbf{u}_{\text{ALE}}), \nabla \psi_j)_{\bar{\Omega}_k},  \\
    \hat{C}^k_{ij} &= C^k_{ij} + (\tau_{\mathbf{u}} \nabla \psi_i, \partial_t \boldsymbol{\phi}_j + \boldsymbol{\phi}_j \cdot \nabla \mathbf{u} + (\mathbf{u} - \mathbf{u}_{\text{ALE}}) \cdot \nabla \boldsymbol{\phi}_j - \nu \nabla^2 \boldsymbol{\phi_j})_{\bar{\Omega}_k}, \\
    \hat{D}^k_{ij} &= D^k_{ij} + (\tau_{\mathbf{u}} \nabla \psi_i, \nabla \psi_j)_{\bar{\Omega}_k},
\end{align}
\noindent where
\begin{align}
    A^k_{ij} &= (\boldsymbol{\phi}_i, \partial_t \boldsymbol{\phi}_j)_{\bar{\Omega}_k} + (\boldsymbol{\phi}_i, \nabla \mathbf{u} \cdot \boldsymbol{\phi}_j)_{\bar{\Omega}_k} + (\boldsymbol{\phi}_i, \mathbf{u} \cdot \nabla \boldsymbol{\phi}_j)_{\bar{\Omega}_k} - (\boldsymbol{\phi}_i, \mathbf{u}_{\text{ALE}} \cdot \nabla \boldsymbol{\phi}_j)_{\bar{\Omega}_k} \nonumber \\
    & \qquad + (\nabla \boldsymbol{\phi}_i, \nu \nabla \boldsymbol{\phi}_j)_{\bar{\Omega}_k} - (\nabla \boldsymbol{\phi}_i, \dfrac{\nu}{2} \llbracket \boldsymbol{\phi}_j \rrbracket \otimes \mathbf{n})_{{\partial \bar{\Omega}}^*_k} - (\boldsymbol{\phi}_i, \nu \{\!\{ \nabla \boldsymbol{\phi}_j \}\!\} \cdot \mathbf{n})_{{\partial \bar{\Omega}}^*_k} \nonumber \\
    & \qquad + (\boldsymbol{\phi}_i, \nu \tau_k \llbracket \boldsymbol{\phi}_j \rrbracket)_{{\partial \bar{\Omega}}^*_k}, \label{Eq-discrete-1} \\
    B^k_{ij} &= - (\nabla \boldsymbol{\phi}_i, \psi_j)_{\bar{\Omega}_k} + (\boldsymbol{\phi}_i, \{\!\{ \psi_j \}\!\} \mathbf{n})_{{\partial \bar{\Omega}}^*_k}, \label{Eq-discrete-2} \\
    C^k_{ij} &= (\psi_i, \nabla \cdot \boldsymbol{\phi}_j)_{\bar{\Omega}_k} - (\psi_i, \dfrac{1}{2} \llbracket \boldsymbol{\phi}_j \rrbracket \cdot \mathbf{n})_{{\partial \bar{\Omega}}^*_k}, \label{Eq-discrete-3} \\
    D^k_{ij} &= 0, \label{Eq-discrete-4}
\end{align}
\noindent and $\mathbf{R}_{\mathbf{u}}$, $\mathbf{R}_{p}$ are the discretized residual terms.

\section{Software implementation}
\label{Sec-Software}

The mortar method is implemented in \texttt{Lethe} \cite{Lethe2025}, an open-source CFD software that is based on the \texttt{deal.II} finite element library \cite{Arndt2025} and uses \texttt{p4est} for mesh management \cite{Burstedde2011}, \texttt{Trilinos} for numerical linear algebra \cite{Heroux2005}, and MPI for parallelization. Although the present work relies on the matrix-free implementation, the MEM implementation is compatible with both matrix-based and matrix-free architectures; more on the implementation and verification of the matrix-free solver can be found in \cite{Saavedra2025}. 

Three main components are used to integrate the mortar method: \texttt{Mortar Manager}, \texttt{Coupling Operator}, and \texttt{Coupling Evaluation}. The \texttt{Mortar Manager} is responsible for storing information regarding mesh alignment, mortar-cell indices,  weights and normal vectors of the quadrature points. The \texttt{Mortar Manager} accounts for the generation of curved mortar cells within the circular geometry of the rotor-stator interface. Such interface is approximated with high-order polynomials and hence the normals are determined exactly. The \texttt{Coupling Operator} stores information corresponding to each interface side ($\partial\bar{\Omega}^+$ and $\partial\bar{\Omega}^-$, as described in Section~\ref{Sec-sub-MortarTerms}), computes the cell penalty factor, and updates the constraints related to the complete system of equations. Lastly, the \texttt{Coupling Evaluation} computes the mortar terms represented in Eq.~\eqref{Eq-NSweakWmortar}.

\subsection{Grid generation and updated configuration} 
\label{Sec-sub-soft-grid}

The grids for the rotor and stator subdomains are generated separately and merged into a unique triangulation. We assume that the number of elements on both sides of the mortar interface is equal, and that such elements are equally sized and spaced. In three-dimensional cases, this restriction is enforced in the planes perpendicular and parallel to the rotation axis. As mentioned in Section~\ref{Sec-GovEqs}, the displacement of the rotor domain mesh (due to an imposed angular velocity) may lead to non-conforming configurations of the rotor-stator interface, and thus, each boundary cell may be connected to up to two mortar cells. Although this assumption of equal angular length of the boundary cells at the mortar interface seems restrictive, it has been a common choice in MEM applications \cite{Ferrer2012, Zhang2015, Durrwachter2021} and could be alleviated in future work by allowing more than two mortar cells to be connected to a single regular cell in either $\bar{\Omega}^+$ or $\bar{\Omega}^-$.

The rotated rotor domain is obtained by a mapping of the initial configuration, using the following relation for the transformation of each node position:
\begin{equation}
    \mathbf{x} = \mathbf{R} \: \mathbf{x}_0,
\end{equation}
\noindent where $\mathbf{x}_0$ is the initial node coordinate, $\mathbf{x}$ is the updated coordinate, and $\mathbf{R}$ is the rotation matrix with respect to the origin. The latter is defined for two-dimensional cases as:
\begin{equation}
    \mathbf{R} \coloneq
    \begin{bmatrix}
        \cos [\theta(t)] & -\sin[\theta(t)] \\
        \sin [\theta(t)] & \cos [\theta(t)]
    \end{bmatrix};
\end{equation}
\noindent in three dimensions, $\mathbf{R}$ is given by:
\begin{equation}
    \mathbf{R} \coloneq \cos[\theta(t)] \: \mathbf{I} + \sin[\theta(t)] \: \mathbf{W} + (1-\cos[\theta (t)]) \: \mathbf{r} \otimes \mathbf{r}
\end{equation}
\noindent where $\mathbf{r}$ is the rotation axis vector around the origin, $\mathbf{I}$ is the identity tensor, and $\mathbf{W}$ is the skew symmetric tensor of $\mathbf{r}$.

\subsection{Creation of mortar operators} 
\label{Sec-sub-soft-operators}

Let us consider the following boundary integral computed at mortar cells, as indicated in Eq.~\eqref{Eq-discrete-1}:
\begin{equation}
    (\boldsymbol{\phi}_i, \nu \tau_k \llbracket \boldsymbol{\phi}_j \rrbracket)_{{\partial \bar{\Omega}}^*_k} \approx \sum_{q=1}^{n_\text{q}} \left( \boldsymbol{\phi}_i (q) \: \nu \tau_k \llbracket \boldsymbol{\phi}_j (q) \rrbracket \right) \: w_q (\det \mathbf{J}_k),
\end{equation}
\noindent where we approximate the integral with a summation over Gauss quadrature points $q$ ($q = 1, \ldots, n_q$); $w_q$ is the weight of each quadrature point, and $\mathbf{J}_k$ is the Jacobian matrix of the transformation from the Cartesian coordinates $\mathbf{x}$ to the mapped reference system $\boldsymbol{\xi}$. 

As indicated in Figure~\ref{Fig-mortar}, the quadrature  and weights at the mortar cells are obtained by mapping a reference quadrature into each mortar segment. The determinant of the Jacobian of this transformation is computed as:
\begin{equation}
    \det \mathbf{J}_k = R \delta_0 \delta_1,  
\label{Eq-Jacobian}
\end{equation}
\noindent in which $R$ is the radius of the in-plane rotor-stator interface, $\delta_0$ is the angle corresponding to the mortar segment at the plane perpendicular to the rotation axis, given by:
\begin{equation}
    \delta_0 = \begin{cases}
    \dfrac{2 \pi R}{n_R}, & \text{for aligned meshes}, \\
    \alpha \:\: \text{or} \:\: \dfrac{2 \pi R}{n_R} - \alpha , & \text{for misaligned meshes},
     \end{cases}
\end{equation}
\noindent where $\alpha$ is the angle between two vertices in the rotor and stator sides (see misalignment example in Figure~\ref{Fig-mortar}), and $n_R$ is the in-plane number of subdivisions.  In \eqref{Eq-Jacobian}, $\delta_1$ is the cell height along the rotation axis, given by:
\begin{equation}
    \delta_1 = \dfrac{H}{n_H},
\end{equation}
\noindent where $H$ is the out-of-plane dimension and $n_H$ is the corresponding number of subdivisions. Since we assume that the element size is the same in the in-plane and out-of-plane configurations, the Jacobian of the transformation is constant. 

The integration is performed in two steps. First, we evaluate the variables at the mortar quadrature points and store the result. For instance, to compute the jump values $\llbracket \boldsymbol{\phi}_j (q) \rrbracket$ we evaluate the $\boldsymbol{\phi}_j (q)$ at both sides of the interface. For the boundary terms containing the normal unit vector $\mathbf{n}$, it is decomposed into $\mathbf{n}^-$ and $\mathbf{n}^+$. Communication is the next step, where data is exported from local ($\bar{\Omega}^-$) to ghost ($\bar{\Omega}^+$) side. After that, we load the stored data and perform the integration of the boundary term. The matrix-free evaluation is performed via the algorithm described in \cite{Heinz2023, Bergbauer2025}, which aims to optimize repeated computations on cached data to adapt to the capabilities of modern hardware. The computation of the remaining boundary integrals at the mortar interface follow the same steps as the one herein described.

\subsection{Mesh parallel decomposition}
\label{Sec-sub-soft-parallel}

\begin{figure}[htb!]
    \centering
    \includegraphics[width=0.95\textwidth]{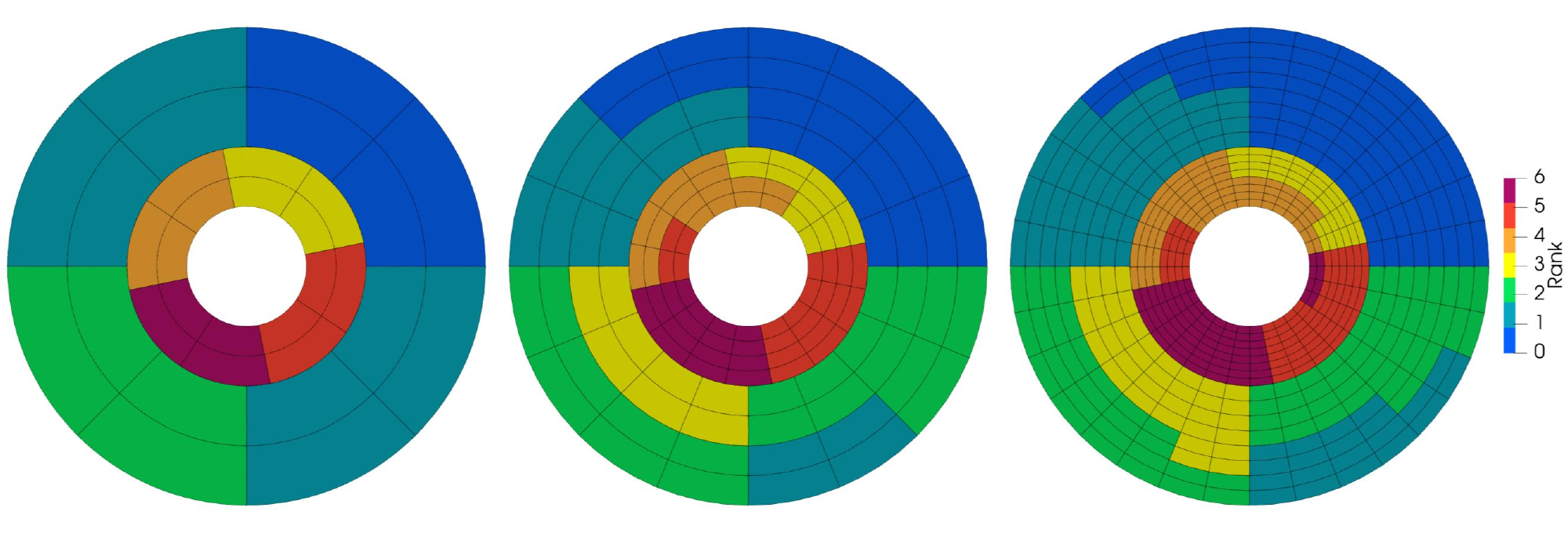}
    \caption{Cell ownership as the mesh is refined; the colors indicate the seven subdomains (ranks), and an iso-parametric mapping is used.}
    \label{Fig-cell-owners-steady}
\end{figure}

\begin{figure}[htb!]
    \centering
    \includegraphics[width=0.95\textwidth]{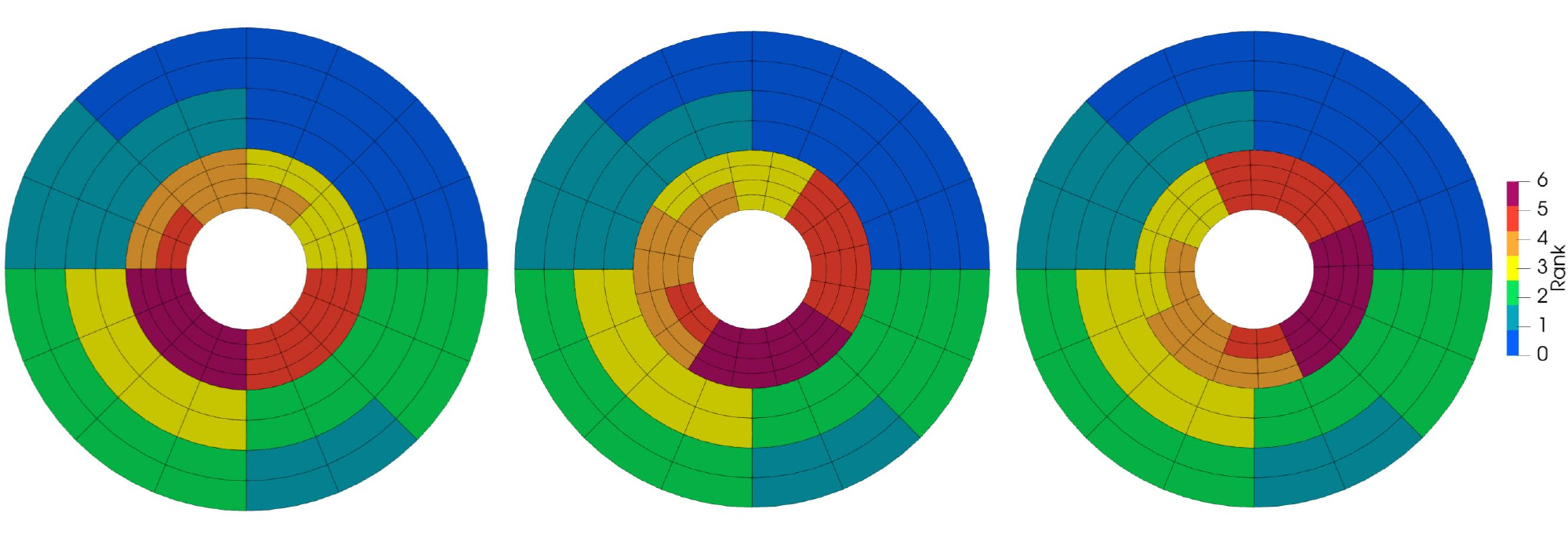}
    \caption{Cell ownership as the rotor mesh is rotated anti-clockwise; the colors indicate the seven subdomains (ranks), and an iso-parametric mapping is used.}
    \label{Fig-cell-owners-transient}
\end{figure}

The mesh parallel decomposition relies on the \texttt{p4est} library \cite{Burstedde2011}. It partitions space-fitting curves, and allows the attribution of different cell weights to be taken into account during load balancing. In the current application, the space-fitting curve is the composition of the stator and rotor sides of the mortar interface. \textbf{Figure~\ref{Fig-cell-owners-steady}} depicts an example of domain partitioning over processes during mesh refinement of a steady case, while \textbf{Figure~\ref{Fig-cell-owners-transient}} shows the partitioning as the rotor domain mapping is rotated. We do not attempt to keep the rotor-stator interface cells within a single process. Child cells do not necessarily remain on the same process as their original unrefined cell, while rotated cells are still owned by the same initial rank.

\subsection{Geometric multigrid preconditioner}
\label{Sec-sub-soft-multigrid}

The GCMG preconditioner builds a hierarchy of discretizations through
geometric coarsening of the mesh ($h$) and reduction of the polynomial degree ($\mathcal{P}$), applying a smoother on each level and a coarse-grid solver on the coarsest one. Available coarse-grid solvers in \texttt{Lethe} include a direct solver, an algebraic multigrid (AMG) solver, an incomplete LU factorization (ILU), and an iterative generalized minimal residual method (GMRES), preconditioned by AMG or ILU. As smoother, we use an inverse diagonal (ID) or the Additive Schwarz method (ASM) accelerated by a relaxation scheme \cite{Saavedra2025}. The GCMG preconditioner does not require global matrices to be assembled, and hence it is compatible with the matrix-free architecture. More details on the GCMG \texttt{deal.II} architecture can be found in \cite{Munch2023}, while its implementation in \texttt{Lethe} is further described in \cite{Saavedra2025}.

The contributions from the mortar interface integrals need to be accounted on each level, in particular, in the smoother and coarse-grid solver. To this end, the relevant element matrices (or their diagonals) are reconstructed column-by-column, and hence the storage of full matrices is avoided -- a common approach in matrix-free algorithms. The corresponding pseudocodes are presented in Algorithms~\ref{Alg-diagonal} and \ref{Alg-matrix}.

\begin{algorithm}[htb!]
\caption{Reconstruction of mortar contribution to matrix diagonal. The loop over cells $c_m$ refers to the cells located at either side of the mortar interface.}
\label{Alg-diagonal}
\begin{algorithmic}
\For{\textbf{each} cell $c_m$}
\For{\textbf{each} local DoF $i$}
\State Create basis vector $e_i$ associated with the shape function $N_i$
\State Evaluate $e_i$ at quadrature points
\State Integrate mortar contributions
\State $A_{c_m}(i,i) \gets i$-th entry of resulting vector
\EndFor
\State Assemble global diagonal
\EndFor
\end{algorithmic}
\end{algorithm}

\begin{algorithm}[htb!]
\caption{Reconstruction of mortar contribution to matrix $\mathbf{A}$. The loop over cells $c_m$ refers to the cells located at either side of the mortar interface.}
\label{Alg-matrix}
\begin{algorithmic}
\For{\textbf{each} cell $c_m$}
\For{\textbf{each} local DoF $i$}
    \State Create basis vector $e_i$ associated with the shape function $N_i$
    \State Evaluate $e_i$ at quadrature points
\EndFor
\EndFor
\State Communicate evaluated values to ghost cells
\For{\textbf{each} cell $c_m$}
\For{\textbf{each} relevant DoF $i$}
    \State Integrate mortar contributions
    \For{\textbf{each} local DoF $j$}
        \State $A_{c_m}(j, i) \gets j$-th entry of resulting vector
    \EndFor
\EndFor
\State Assemble global matrix
\EndFor
\end{algorithmic}
\end{algorithm}

\section{Numerical examples}
\label{Sec-Examples}

In this section we present study cases to verify and validate the MEM implementation within \texttt{Lethe}. A steady-state two-dimensional example is described in Section~\ref{Sec-sub-steadyMMS}, where a manufactured solution is employed so that the convergence rate of high-order polynomial approximations can be observed. In Section~\ref{Sec-sub-transientTGV}, a transient two-dimensional Taylor-Green vortex example is used to verify the convergence order of viscid and inviscid flows. In both two-dimensional cases, we study the change in error norms due to different mesh alignment patterns at the interface. The performance of the solver is assessed in Section~\ref{Sec-sub-performance}, where a three-dimensional example is used to show strong and weak scalability of the model. Following that, Sections~\ref{Sec-sub-2Drushton} and~\ref{Sec-sub-3Dpbt} showcases simulations of stirred mixing tanks in two- and three-dimensions, respectively, to investigate mesh convergence and validate the model with existing numerical and experimental data.

In the following examples, unless otherwise stated, we solve the linearized Navier-Stokes equations using the GMRES method, a GCMG preconditioner \cite{Wesseling2001, Munch2023} and an inverse diagonal smoother as a relaxation scheme. In the transient examples, a second-order implicit backward-difference time integration scheme (BDF2) is employed. The following examples employ equal high-order elements $Q_\mathcal{P}Q_\mathcal{P}$ ($\mathcal{P} \geq 2$, where $\mathcal{P}$ is the polynomial order) for the velocity and pressure fields. While the SUPG-PSPG is a robust stabilization for the CG framework of the Navier-Stokes solver, the introduction of the DG mortar interface generated pressure instability for the linear polynomial approximation $Q_1Q_1$. Additional tests have shown that adding a bubble-type enrichment function \cite{Zienkiewicz1977} to the velocity field recovers the interface continuity in the matrix-based solver. Numerical examples considering enriched $Q_1Q_1$ elements, as well as a discussion on alternative stabilization methods for the linear approximation, will be addressed in future work.

\subsection{Steady-state manufactured solution}
\label{Sec-sub-steadyMMS}

\begin{figure}[htb!]
\centering
\begin{subfigure}[b]{0.45\textwidth}
\centering
    \includegraphics[scale=1]{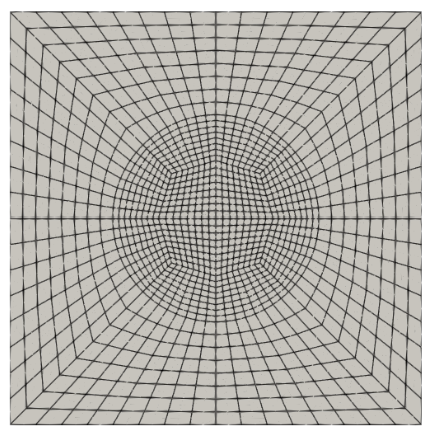}
\caption{Mesh configuration (refinement level $l=4$) used in the two-dimensional MMS simulation (Section~\ref{Sec-sub-steadyMMS}) and TGV case (Section~\ref{Sec-sub-transientTGV}). The corresponding domain is $\Omega = [-1, 1] \times [-1, 1]$ and the rotor-stator interface is placed at a radial distance of $0.5$ from the center point.}
\label{Fig-mms2d-mesh}
\end{subfigure}
\hfill
\begin{subfigure}[b]{0.45\textwidth}
\centering
    \includegraphics[scale=0.25]{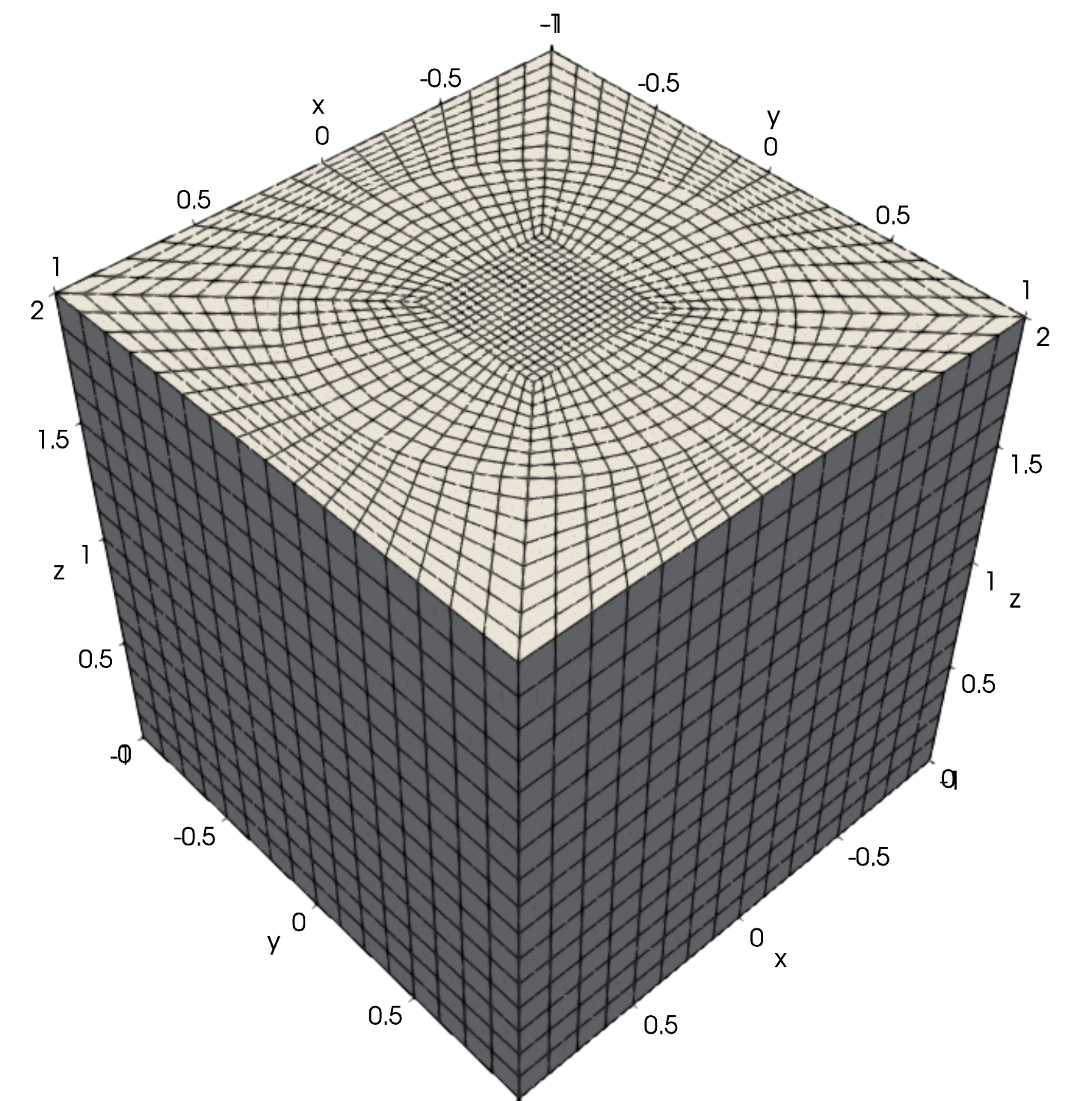}
\caption{Mesh configuration (refinement level $l=2$) used in three-dimensional MMS scalability study (Section~\ref{Sec-sub-performance}). The corresponding domain is $\Omega = [-1, 1] \times [-1, 1] \times [0, 2]$ and the cylindrical mortar interface is located at radial distance of $0.5$ from the rotation axis $z$.}
\label{Fig-mms3d-mesh}
\end{subfigure}
\caption{Mesh discretization for (a) two-dimensional and (b) three-dimensional manufactured solution examples.}
\end{figure}

We verify the MEM implementation by simulating a two-dimensional steady-state problem with the Method of Manufactured Solutions (MMS). The manufactured problem is:
\begin{align}
    u &= -2\sin(\pi x)^2 \cos(\pi y) \sin(\pi y), \\
    v &= 2 \cos(\pi x) \sin(\pi x) \sin(\pi y)^2, \\
    p^* &= \sin(\pi x) \sin(\pi y),
\end{align}
\noindent where $u, v$ are the $x$ and $y$ components of the velocity vector $\mathbf{u}$. The problem domain is a combination of a plate with a hole and a circle, which yields a simple merged geometry that still allows us to evaluate the MEM accuracy (see \textbf{Figure~\ref{Fig-mms2d-mesh}}). The combined rotor-stator domain is defined in $[-1,1] \times [-1,1]$, with the circular interface located at a radial distance of $0.5$ from the center point. In this example, no rotation velocity is imposed on the rotor domain ($\omega=0$). 

\begin{figure}[htb!]
\centering
    \includegraphics[width=0.9\textwidth]{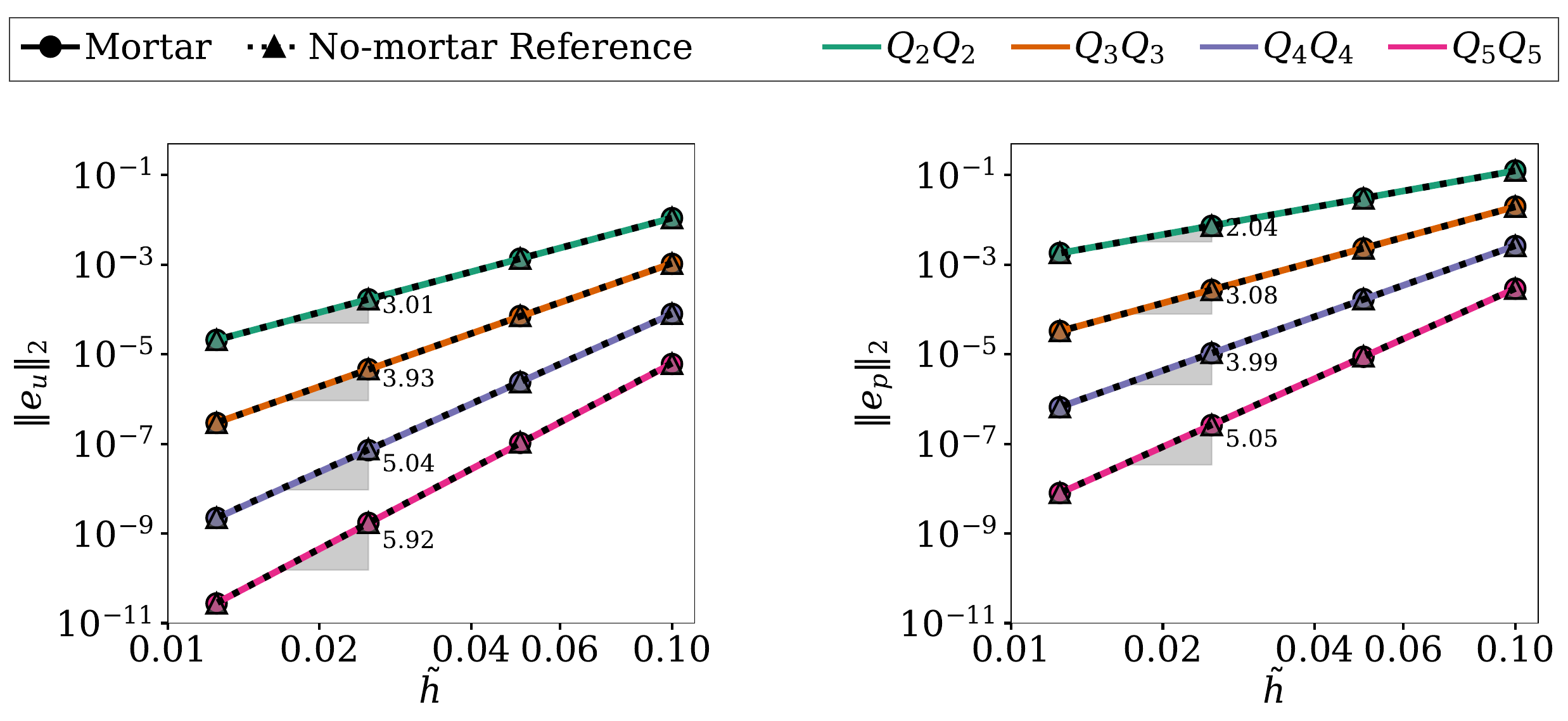}
\caption{$L_2$ norm of the error of velocity (left) and pressure (right) fields for the two-dimensional MMS case. In the reference case, the mesh in Figure~\ref{Fig-mms2d-mesh} is used without a mortar interface. For the rotor-stator mesh, no rotor rotation was imposed.}
\label{Fig-convMMS}
\end{figure}

The simulation is carried out for a kinematic viscosity of $\nu = 1$, four refinement levels (the refinement level $l=1$ contains $80$ elements), and polynomial approximations from $\mathcal{P} = 2$ to $\mathcal{P} = 5$. In this steady case, the time-dependent term indicated in Eq.~\eqref{EqNS} is neglected. \textbf{Figure~\ref{Fig-convMMS}} depicts the $L_2$ error norms, where we notice the optimal $\mathcal{O}(h^{\mathcal{P}+1})$, $\mathcal{O}(h^\mathcal{P})$ rates for velocity and pressure fields, respectively. Since the rotor-stator mesh is non-uniform, the reported values of $\tilde h$ refer to a characteristic mesh size based on the average element area. The $L_2$ norms of the error are compared to reference values (indicated as (R) in Figure~\ref{Fig-convMMS}), in which the same parameters and mesh are used to simulate the MMS case without a mortar interface. Said error norms are identical for the subdivided rotor-stator domain and for the reference case.

\begin{figure}[htb!]
    \centering
    \includegraphics[width=0.9\textwidth]{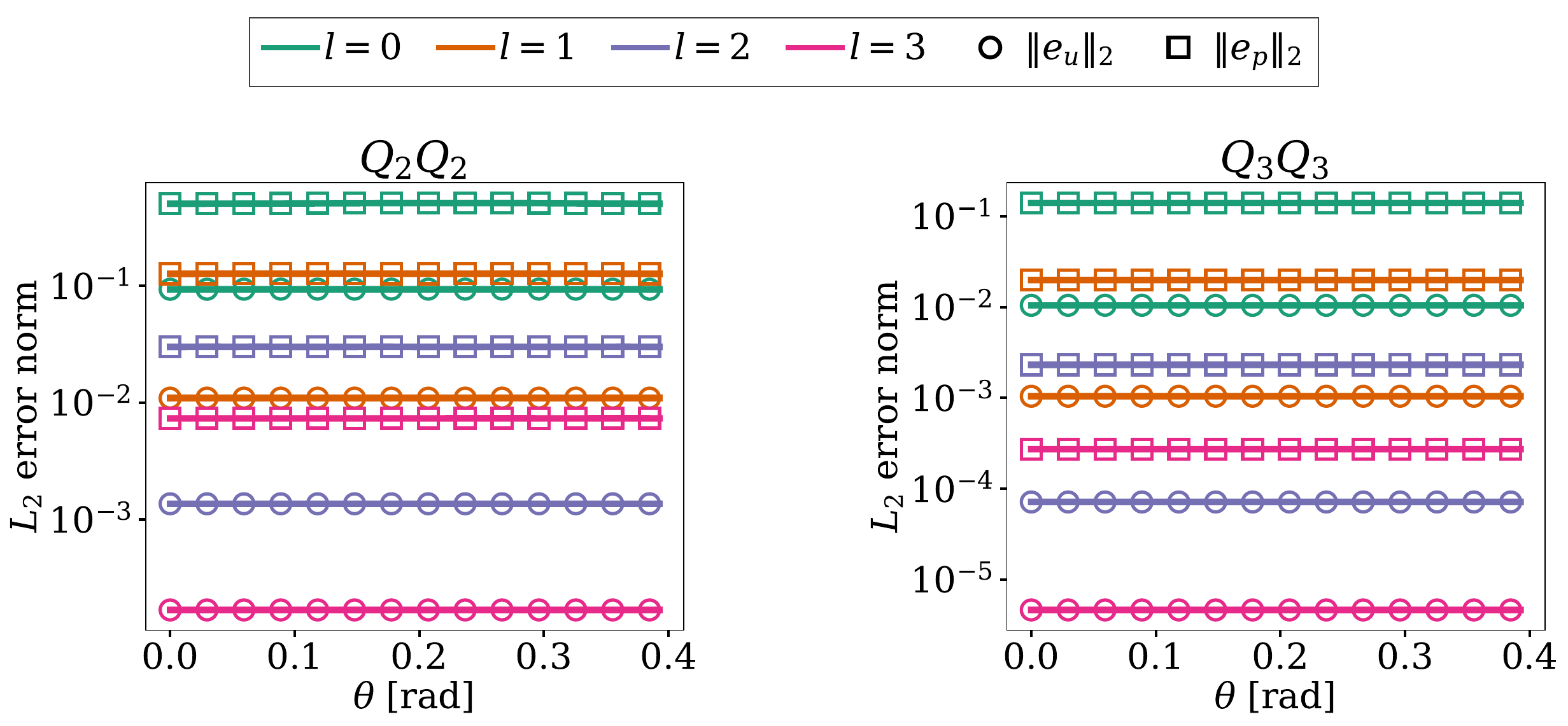}
\caption{$L_2$ error norm for the two-dimensional MMS problem considering different initial rotation angles of the rotor mesh. The points in each curve are taken from the interval $[0, \theta_{max}]$, where $\theta_{max} = \pi/8$ is the cell angular length at the mortar interface for the coarsest refinement level.}
\label{Fig-rot-vs-angle-p}
\end{figure}

Considering the same problem, we analyze the dependence of the error on the rotor mesh rotation. The goal is to verify whether the alignment of the mortar cells with respect to the stator boundary cells (see illustration in Figure~\ref{Fig-mortar}) interferes in the accuracy of the solution. Considering that the angular length of each interface cell in the coarsest mesh is $\theta_{max} = 0.3926$~rad, we take $200$ sampling values in the interval $[0, \theta_{max}]$ to verify the corresponding error norms. \textbf{Figure~\ref{Fig-rot-vs-angle-p}} depicts the $L_2$ error norm for four refinement levels, considering $Q_2Q_2$ and $Q_3Q_3$ elements. Regardless of the refinement level, the error norms for both velocity and pressure fields remain constant, confirming that the geometric incompatibility generated at the interface due to mortar cell misalignment does not lead to loss of convergence properties. This conclusion is in alignment with the fact that a reference quadrature is mapped into each mortar segment intersected by the boundary cells edges (Figure~\ref{Fig-mortar}), and thus the quadrature order is preserved regardless of the mortar element size. The same behavior has been observed in \cite{Ferrer2012} for a DG formulation. 

\subsection{Taylor-Green vortex}
\label{Sec-sub-transientTGV}

The transient MEM implementation is verified with a two-dimensional Taylor-Green vortex (TGV) case. The same domain geometry and mesh discretization from Section~\ref{Sec-sub-steadyMMS} is employed, as indicated in Figure~\ref{Fig-mms2d-mesh}, and an angular velocity $\omega$ is prescribed in the rotor domain. The initial condition is given by:
\begin{align}
    u &= \cos(x) \sin(y), \\
    v &= -\sin(x) \cos(y), \\
    p &= -\dfrac{1}{4} [\cos(2x) + \cos(2y)],
\end{align}
\noindent in which $u,v$ are the $x$ and $y$ components of the velocity vector $\mathbf{u}$, respectively. The simulation considers a kinematic viscosity of $\nu=1$ and a time step of $\Delta t = 10^{-4}$~s during $t_{end}=0.1$~s. The corresponding analytical solution is given by:
\begin{align}
    u &= \exp(-2 \nu t) \cos(x) \sin(y), \\
    v &= - \exp(-2 \nu t) \sin(x) \cos(y).
\end{align}

\begin{figure}[htb!]
\centering
    \includegraphics[width=0.9\textwidth]{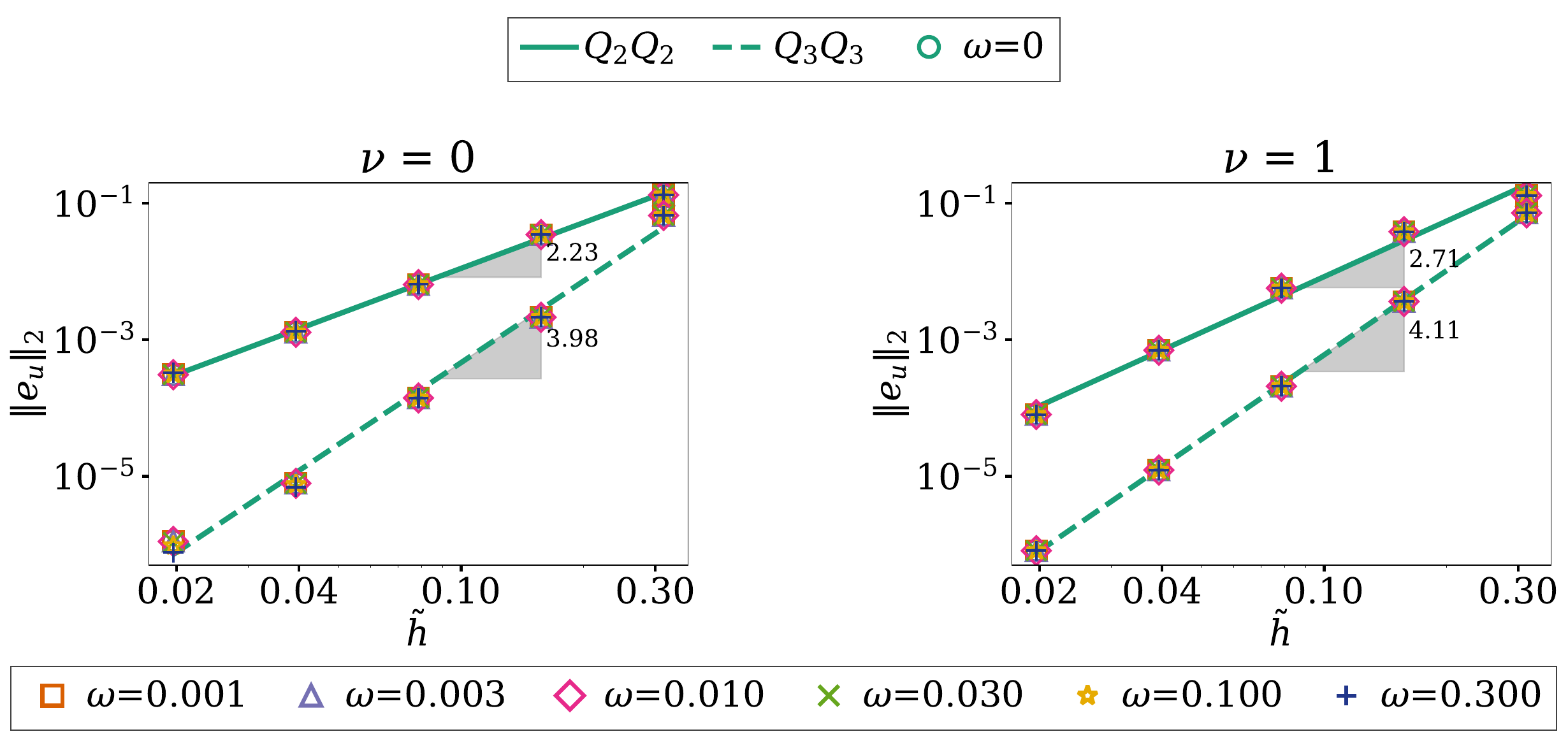}
\caption{$L_2$ error of velocity field in the two-dimensional TGV problem for different angular velocities of the rotor mesh, considering viscid (left, $\nu=1$) and inviscid (right, $\nu=0$) cases. The shaded gray triangles indicate the convergence rate for all angular velocity values, which have units of rad/s.}
\label{Fig-convTGV}
\end{figure}

\textbf{Figure~\ref{Fig-convTGV}} shows velocity $L_2$ error norms for a case with no rotor mesh rotation ($\omega=0$), as well as other randomly chosen values of rotor angular velocity. The expected convergence rate $\mathcal{O}(h^{\mathcal{P}+1})$ is obtained for all $\omega$ in the $Q_3Q_3$ case. For $\mathcal{P}=2$, convergence is deteriorated due to the odd-degree structure of the velocity Taylor expansion. Additional tests showed that the same pattern is observed for fourth-order polynomial shape functions and that fifth-order polynomials ($Q_5Q_5$) recovered the expected order $\mathcal{O}(h^{\mathcal{P}+1})$. Convergence rates for $\omega=0.001$ to $\omega=0.3$ rad/s are almost identical to the case with no mesh rotation, which can be noticed as the error markers overlap. The difference in error norms spans from zero to $10^{-4}$, decreasing with element size. 

To verify the accuracy of the scheme in the absence of viscous dissipation effects, we repeated the convergence analysis for an inviscid case ($Re \rightarrow \infty$), assuming $\nu = 0$. Figure~\ref{Fig-convTGV} depicts the convergence rates for the velocity field considering the same rotating speeds as in the viscid case. The difference in error norms also stays in the order of $10^{-4}$ when we compare $\omega \neq0$ to the stationary case, and we observe the same convergence trends for $\mathcal{P}=2, 3$ from the viscid case.

\begin{figure}[htb!]
\centering
    \includegraphics[width=0.9\textwidth]{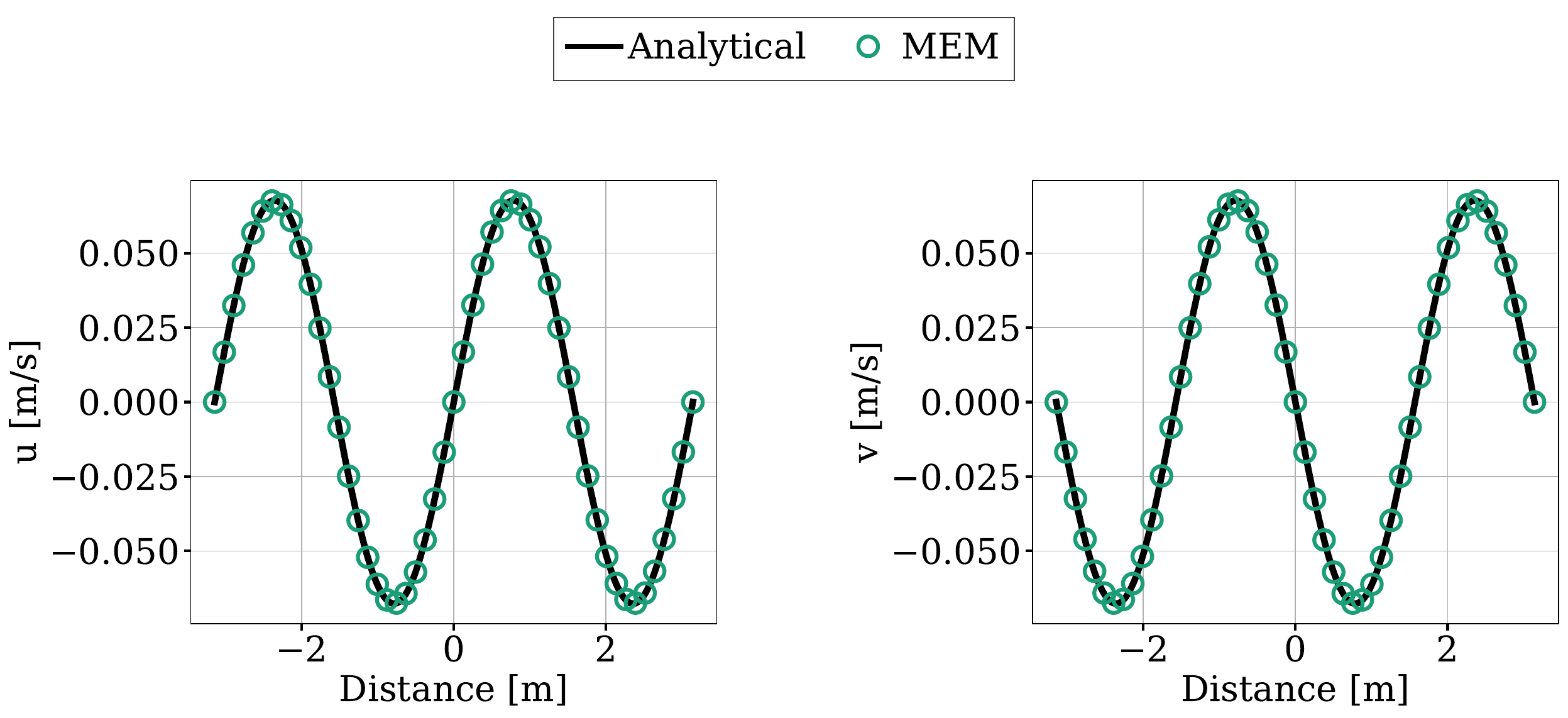}
\caption{Comparison between analytical and numerical solution for two-dimensional TGV case in a diagonal cross-section from ($x=-1, y=-1$) to ($x=1, y=1$).}
\label{Fig-anTGV}
\end{figure}

Considering a diagonal cross-section from $(-1, -1)$ to $(1, 1)$, we compare the numerical results with the analytical solution as depicted in \textbf{Figure~\ref{Fig-anTGV}}, and good agreement is achieved. In this case, the finest refinement level reported in the convergence study (Figure~\ref{Fig-convTGV}), i.e. $5120$ elements, is employed. The computed absolute errors vary between $10^{-4}$ and $10^{-13}$, and higher values are reported in the stator mesh, where the average element size is larger than in the rotor mesh.

\subsection{Performance and scalability tests}
\label{Sec-sub-performance}

To verify scalability of the matrix-free MEM implementation, we study a three-dimensional MMS case in which the mortar interface is introduced by combining a cylinder and an extruded plate with hole, as depicted in \textbf{Figure~\ref{Fig-mms3d-mesh}}. The manufactured solution is expressed as:
\begin{align}
    u & = 2 \pi^2 [-3 \cos(2\pi x)+2] \sin(\pi y) \sin(\pi z) \cos(\pi y) \cos(\pi z) \nonumber \\
    & \qquad + \pi [2 \cos(\pi y)^2 - \cos(\pi z)^2] \sin(\pi x)^3 \sin(\pi y)^2 \sin(\pi z)^2 \cos(\pi x), \\
    v & = 2 \pi^2 [-3\cos(2\pi y)+2] \sin(\pi x) \sin(\pi z) \cos(\pi x) \cos(\pi z) \nonumber \\
    & \qquad+ \pi [2\cos(\pi x)^2 -\cos(\pi z)^2] \sin(\pi x)^2 \sin(\pi y)^3  \sin(\pi z)^2 \cos(\pi y), \\
    w & = 4 \pi^2 [3 \cos(2 \pi z)-2] \sin(\pi x) \sin(\pi y) \cos(\pi x) \cos(\pi y) \nonumber \\
    & \qquad + 2\pi [\cos(\pi x)^2 + \cos(\pi y)^2] \sin(\pi x)^2 \sin(\pi y)^2 \sin(\pi z)^3 \cos(\pi z), \\
    p & = 0.
\end{align}

The steady-state problem is tested with two mesh configurations with the number of cells at the coarse level equal to $l_0~=~18$ and $l_0~=~40$, for $\mathcal{P}=2$ and $\mathcal{P}=3$. The corresponding number of cells and degrees of freedom (DoFs) are reported in \textbf{Table~\ref{Tab-performance-mesh-data}}. For the $Q_3Q_3$ case, we use an $hp$-multigrid preconditioner so that the polynomial degree of the coarsest grid level is reduced by one order, i.e. $\mathcal{P}=2$ (the limitation on using $\mathcal{P}=1$ is discussed at the beginning of Section \ref{Sec-Examples}). The simulations are run on the Trillium distributed memory cluster of the Digital Research Alliance of Canada \cite{Trillium} and use up to 64 nodes with 192 CPU cores each (12,288 CPU cores in total).

\begin{table}[htb!]
\centering
\caption{Number of cells and degrees of freedom (DoFs) for each refinement level employed in the scalability study of Section \ref{Sec-sub-performance}.}
\label{Tab-performance-mesh-data}
\begin{tabular}{cccccc}
    \hline        
    $l_0$   & $l$   & Cells     & DoFs ($Q_2 Q_2$)  & DoFs ($Q_3 Q_3$)  \\ \hline
            & 4     & $164$k    & $5.4$M            & $18$M             \\
    40      & 5     & $1.3$M    & $43$M             & $143$M            \\
            & 6     & $10.5$M   & $338$M            & $1.1$B            \\
            & \\
    18      & 5     & $590$k    & $19$M             & $64$M             \\
            & 6     & $4.7$M    & $152$M            & $513$M            \\ \hline
\end{tabular}
\end{table}

The average number of linear solver iterations per Newton iteration for each refinement level are reported in \textbf{Table~\ref{Tab-performance-iters}}; in all cases, the solver needed three non-linear Newton iterations.  To evaluate the performance of the smoother preconditioner (i.e., relaxation scheme), results for both inverse diagonal (ID) and Additive Schwarz Method (ASM) options are presented. Furthermore, we consider a reference no-mortar case -- same mesh configuration with a merged rotor-stator interface -- to evaluate the impact that the mortar cells have on the solver convergence. The case with $l_0=40$ yields less iterations overall; this can be related to the fact that such discretization contains cells with a more uniform aspect ratio compared to when $l_0 = 18$. We conclude that the number of linear solver iterations is more affected by the aspect ratio of the cells than the presence of the mortar interface. 

In terms of the smoother preconditioner type, ASM yields about a third less iterations than ID; however, it can take up to 3.4 times more computational time depending on the mesh refinement and the number of computer nodes. This result is expected considering the nature of these two relaxation schemes; while ID has a low cost for setup and application, ASM setup is more expensive while it usually leads to less iterations \cite{Saavedra2025}. When comparing the mortar ($m$) and the no-mortar ($n$) cases, a similar number of iterations is shown for the ASM scheme. Meanwhile, a higher iteration count for the $n$ case is noticed for the ID scheme when $l_0 = 18$ when compared to the mortar results.

\begin{figure}[htb!]
\centering
    \includegraphics[width=0.9\textwidth]{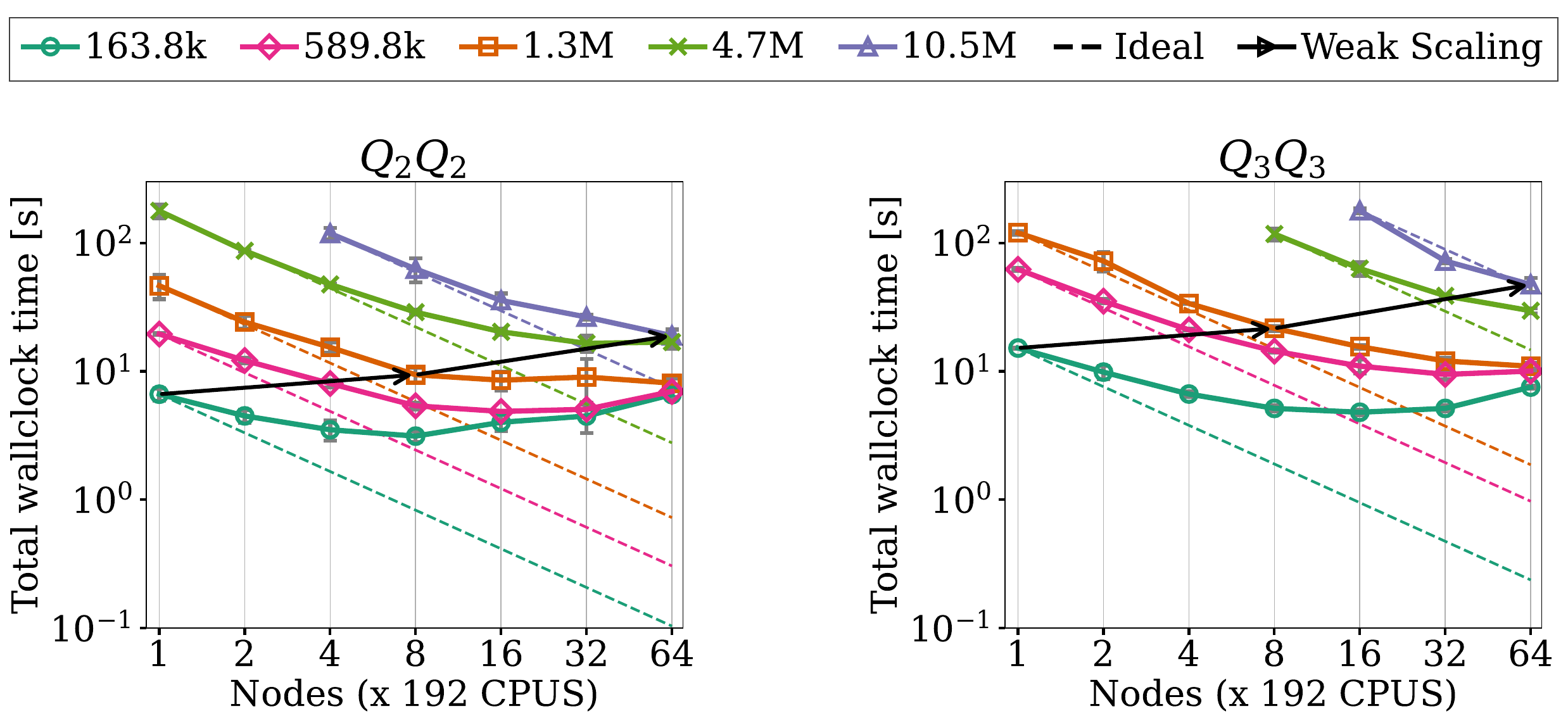}
\caption{Scalability study for a three-dimensional MMS problem considering the total wallclock time. The ideal strong scaling for each refinement level is indicated by dashed lines, while the black arrows indicate the weak scaling for the current case.}
\label{Fig-mms3d-time-total}
\end{figure}

\textbf{Figure~\ref{Fig-mms3d-time-total}} depicts the total wallclock time for the simulated cases. The strong scaling for the coarser meshes shows poorer results since very few cells are contained within each core, and hence communication significantly affects the wallclock time. Furthermore, the solution of a steady case can be harder compared to transient cases, since the flow needs to be solved at once. Significant improvement can be noticed for the two finest meshes, where the $Q_3Q_3$ curve shows an almost ideal strong scaling. The black arrows in the figure indicate the weak scalability, showing an increase in simulation time when both the mesh and the number of nodes are increased by a factor of $8$. From $163.8$k to $1.3$M cells this increase is $\sim 41 \%$, while from $1.3$M to $10.5$M cells is $\sim 105 \%$. We highlight that weak scaling is also affected by the mismatch between the coarse meshes and the elevated number of nodes, where the communication cost predominates.

\begin{table}[htb!]
\centering
\caption{Average number of linear iterations per Newton iteration for each refinement level obtained in the scalability study of Section \ref{Sec-sub-performance}. Results are reported for two relaxation schemes: inverse diagonal (ID) and Additive Schwarz Method (ASM). For each scheme, the rotor-stator results (indicated by $m$) are compared to a reference no-mortar case (indicated by $n$) in which the rotor-stator interface is merged -- leading to the same number of cells but less DoFs.}
\label{Tab-performance-iters}
\begin{tabular}{cccccccccccccc}
    \hline        
    & & & \multicolumn{5}{c}{$Q_2Q_2$} & & \multicolumn{5}{c}{$Q_3Q_3$} \\ \cline{4-8}\cline{10-14} 
    $l_0$ & $l$ & Cells       & \multicolumn{2}{c}{\textbf{ID}} & & \multicolumn{2}{c}{\textbf{ASM}} & & \multicolumn{2}{c}{\textbf{ID}} & & \multicolumn{2}{c}{\textbf{ASM}} \\
    & & & $m$ & $n$ & & $m$ & $n$ & & $m$ & $n$ & & $m$ & $n$ \\ \hline
    & 4 & $164$k      & 21  & 17 & & 7   & 5   & & 18 & 19 & & 8   & 8 \\
    $40$ & 5 & $1.3$M      & 23  & 19 & & 7   & 6   & & 20 & 20 & & 8   & 7 \\
    & 6 & $10.5$M     & 25  & 21 & & 7   & 5   & & 21 & 21 & & 7   & 7 \\
    & \\
    $18$ & 5 & $590$k      & 25  & 33 & & 10  & 10  & & 27 & 46 & & 11  & 15 \\
    & 6 & $4.7$M      & 29  & 50 & & 10  & 11  & & 30 & 63 & & 11  & 14 \\ \hline
\end{tabular}
\end{table}

\begin{figure}[htb!]
\centering
    \includegraphics[width=0.9\textwidth]{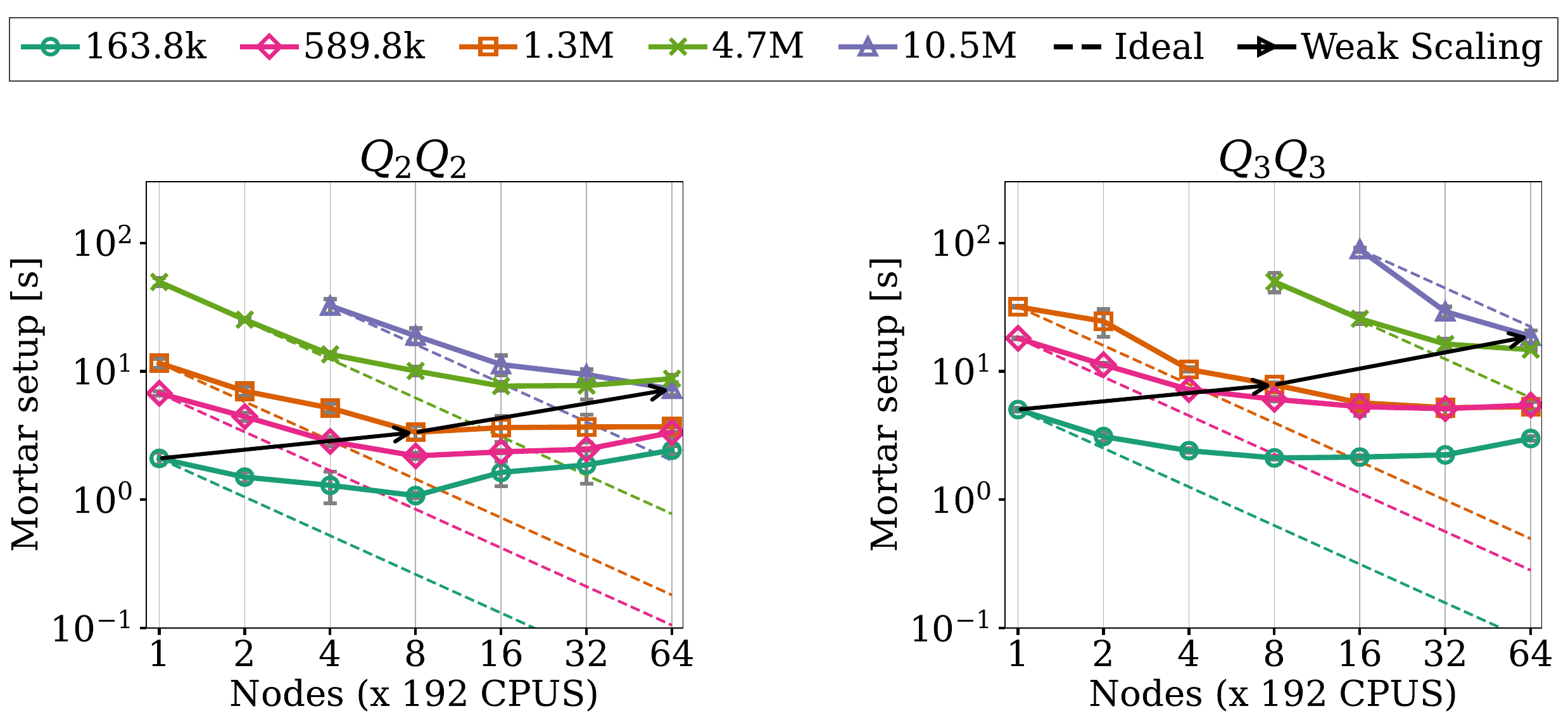}
\caption{Scalability study for a three-dimensional MMS problem considering the mortar setup time. The ideal strong scaling for each refinement level is indicated by dashed lines, while the black arrows indicate the weak scaling for the current case.}
\label{Fig-mms3d-time-mortar}
\end{figure}

As noted in \textbf{Figure~\ref{Fig-mms3d-time-mortar}}, the mortar setup follows the same scalability trend as the total wallclock time. The reported values include time spent in creation of mortar operators, computation of boundary integrals, and update of the rotor mapping. Although the overall cells are distributed among processes considering load balancing, we assume that the cells at the rotor-stator interface and the remaining cells have the same weight. A question that arises is whether the load imbalance of mortar cells could have an effect on the efficiency of the solver. For such analysis, we first define an ideal workload as:
\begin{equation}
    L_\text{ideal} = \dfrac{\sum_i^{n_\text{p}} m_i}{n_\text{p}},
\end{equation}
\noindent in which $n_\text{p}$ is the total number of processes and $m_i$ is the number of mortar cells at each process $i$. The actual workload imbalance is computed as the ratio between the maximum number of cells in a process $m_\text{max}$ and the ideal workload:
\begin{equation}
    L = \dfrac{m_\text{max}}{L_\text{ideal}}.
\end{equation}

\begin{figure}[htb!]
\centering
    \includegraphics[width=0.9\textwidth]{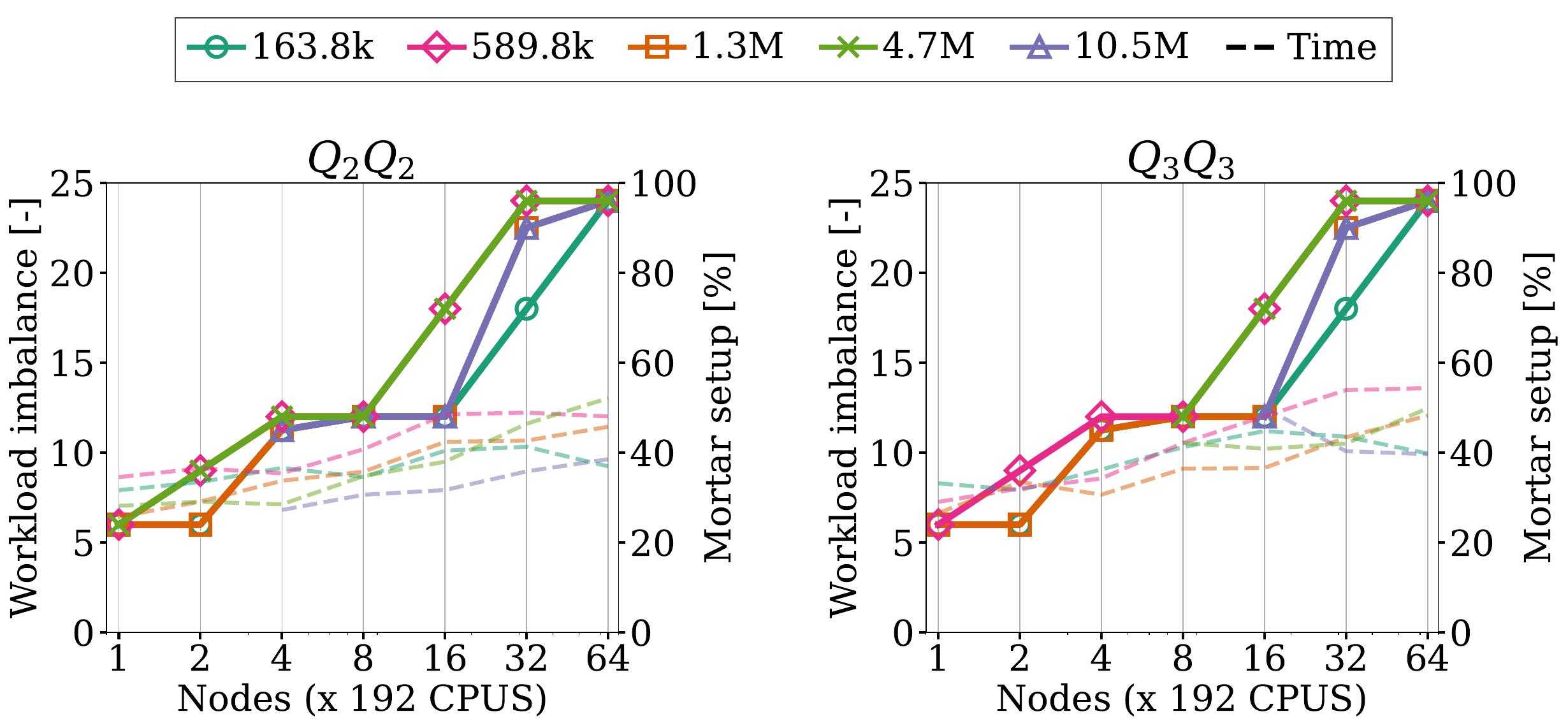}
\caption{Mortar workload imbalance for a three-dimensional MMS problem. The solid lines represent the workload imbalance reported in left $y$-axis, while the dashed lines indicate the percentage of mortar setup time with respect to the total simulation time and are reported in the right $y$-axis.}
\label{Fig-mms3d-workload}
\end{figure}

The computed workload imbalance is reported in \textbf{Figure~\ref{Fig-mms3d-workload}}, as well as the percentage of time spent within the mortar setup. The latter varies around $20$ to $50 \%$ of the total time, significantly contributing to the total wallclock time. Since different weights are not attributed to mortar cells during load balancing, workload imbalance increases significantly with the number of processes. Its deviation from the ideal ratio of $1$ can be improved if we increase the weight of mortar cells, which consists in our next verification. For that, we vary the ratio between the mortar cell weight $w_m$ and the regular cell weight $w_r$, monitoring the mortar workload imbalance factor.

\begin{figure}[htb!]
\centering
    \includegraphics[width=0.9\textwidth]{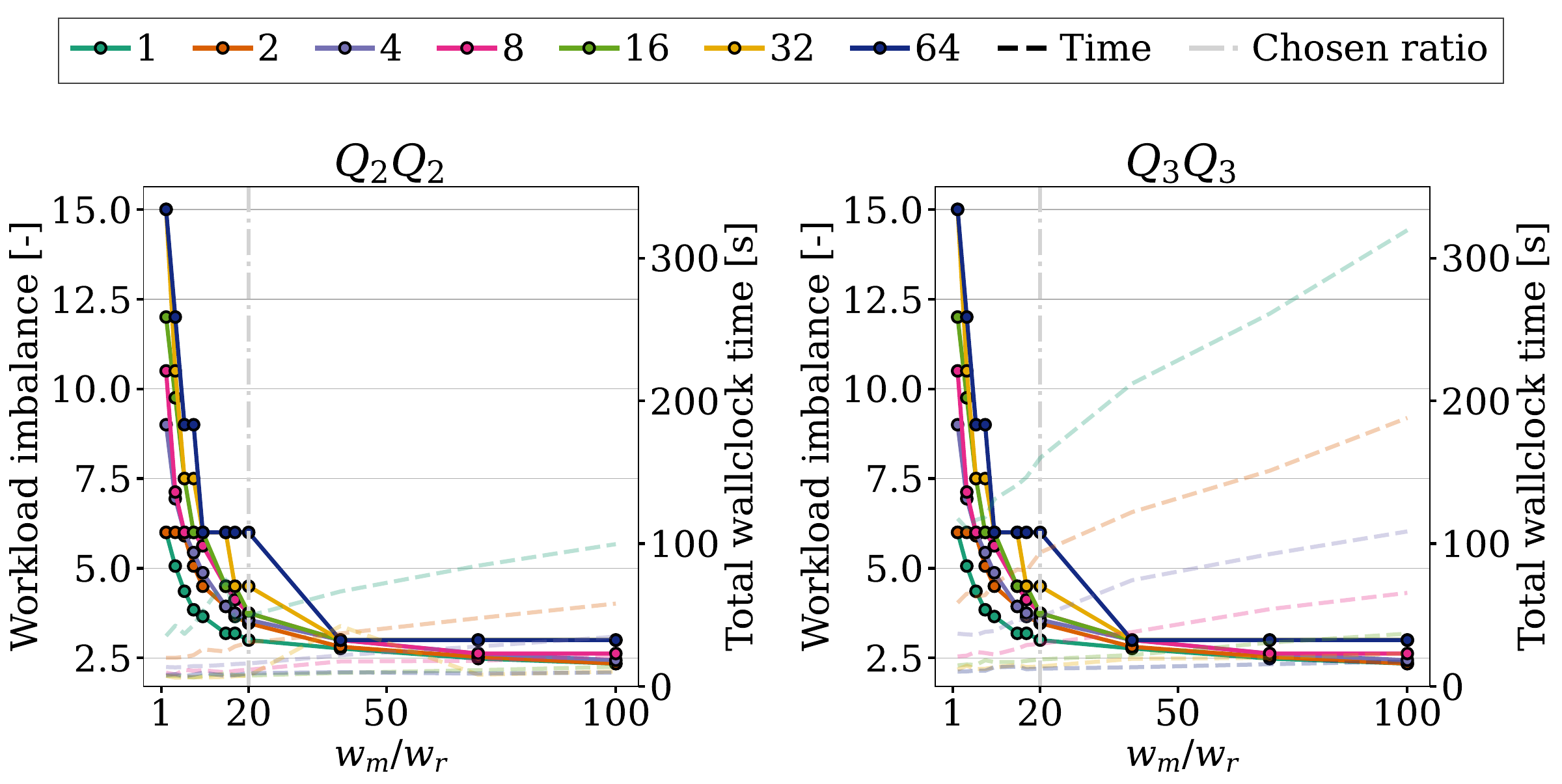}
\caption{Mortar workload imbalance for a three-dimensional MMS problem with $1.3$M cells and various number of computing nodes. The solid lines correspond to the left $y$-axis, and the dashed lines represent the total wallclock time reported in the right $y$-axis. Each node correspond to 192 CPU cores. The vertical dot-dashed gray line indicates the cell weight ratio that yields lower workload imbalance without significantly increasing the corresponding computing time.}
\label{Fig-mms3d-workload-balancing}
\end{figure}

\begin{figure}[htb!]
\centering
    \includegraphics[width=0.9\textwidth]{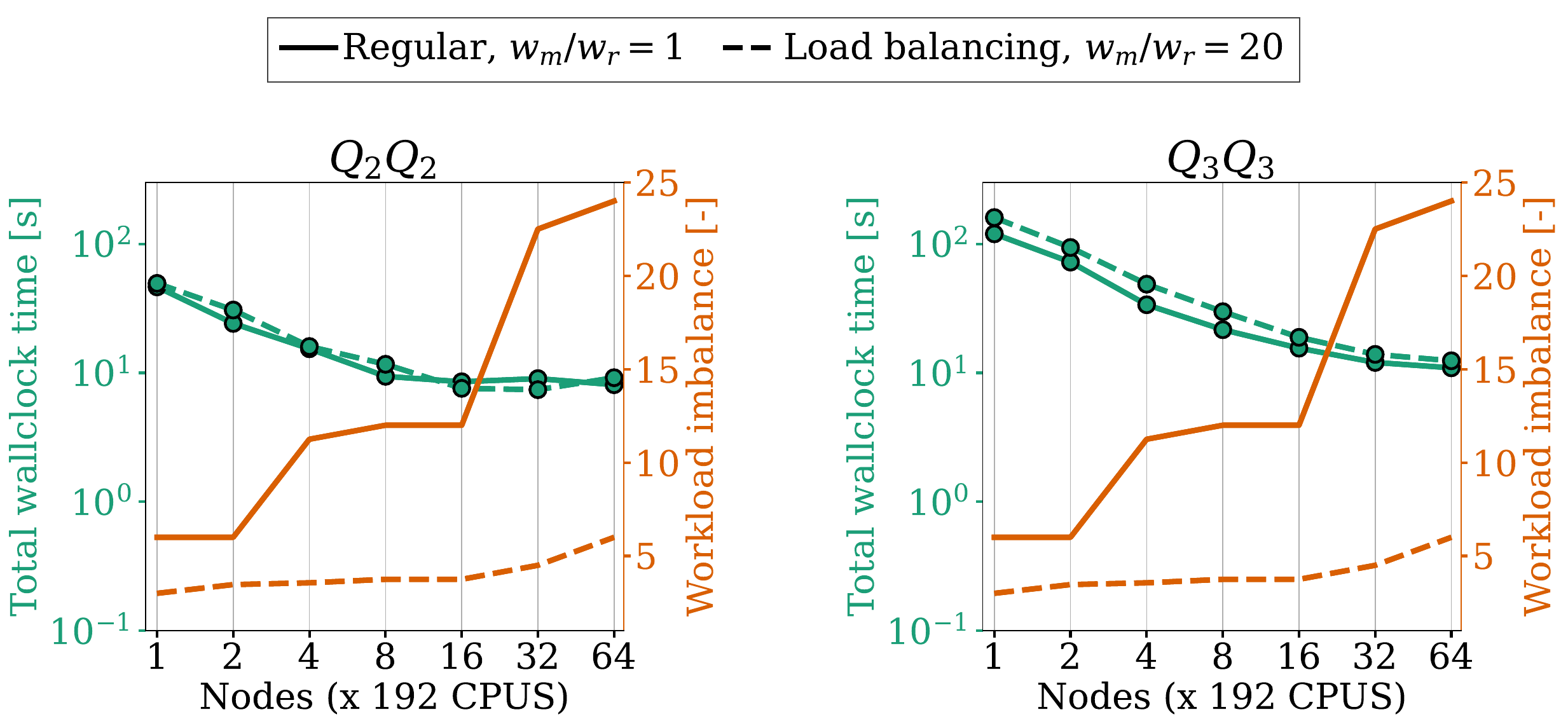}
\caption{Comparison of the total wallclock time (left $y$-axis) and the mortar workload imbalance (right $y$-axis) for (i) the regular case where no additional load balancing is performed and (ii) the case where load balancing is performed considering a cell weight ratio of $w_m/w_r = 20$.}
\label{Fig-mms3d-time-total-compare}
\end{figure}

For the case of $1.3$M cells, \textbf{Figure~\ref{Fig-mms3d-workload-balancing}} depicts the mortar workload imbalance for various numbers of nodes (each node corresponds to 192 CPU cores). The workload stabilizes at approximately $2.5$ when $w_m/w_r >50$; to analyze the impact that the mortar load balancing has on the simulation time, we choose $w_m/w_r = 20$. When looking at the total wallclock time, especially in the $Q_3Q_3$ case, the cell weight ratio of $20$ yields a good balance between improving the load imbalance without significantly increasing the computing time. \textbf{Figure~\ref{Fig-mms3d-time-total-compare}} compares the total wallclock time and the mortar workload imbalance for $w_m/w_r = 1$ and $w_m/w_r = 20$. Even though the load imbalance is significantly reduced -- up to five times less, small variations are observed in the simulation time. In fact, performing the workload balancing requires slightly more time, confirming that it is an unnecessary step in the present mortar approach.

\subsection{Two-dimensional Rushton mixer}
\label{Sec-sub-2Drushton}

\begin{figure}[htb!]
\centering
\begin{subfigure}[b]{0.45\textwidth}
\centering
    \includegraphics[scale=0.35]{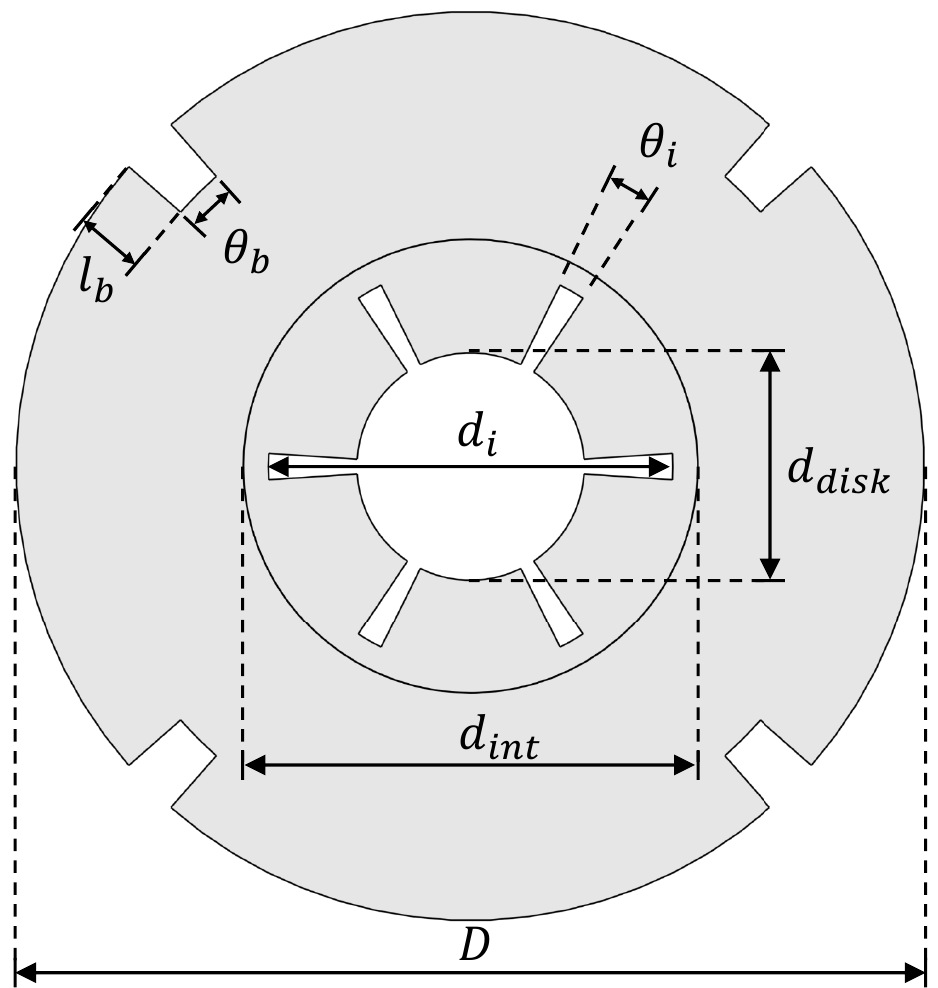}
\caption{Geometry}
\label{Fig-rushton-geometry}
\end{subfigure}
\hfill
\begin{subfigure}[b]{0.45\textwidth}
\centering
    \includegraphics[scale=0.25]{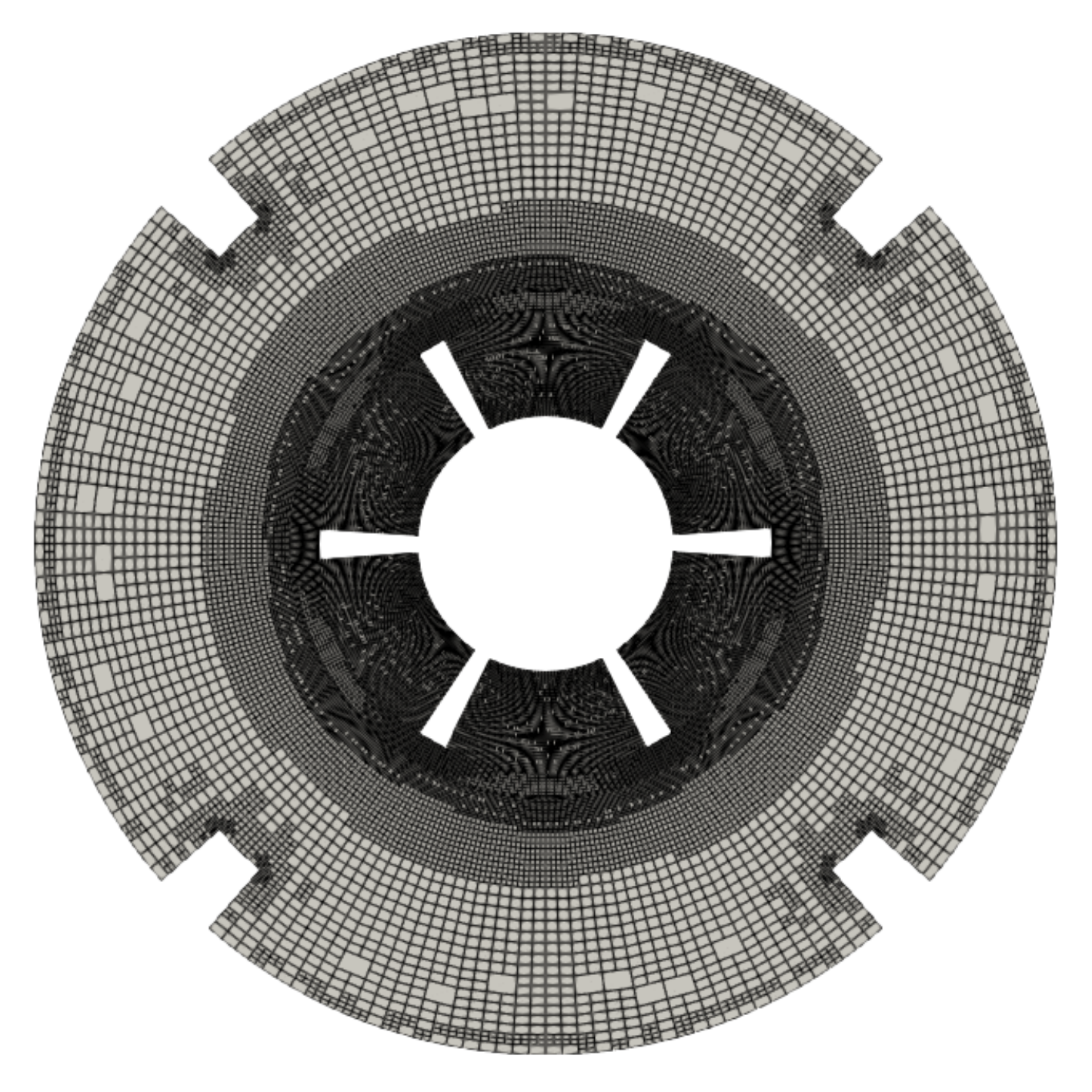}
\caption{Mesh with adaptive refinement}
\label{Fig-rushton-mesh}
\end{subfigure}
\caption{Two-dimensional Rushton mixer: (a) geometry and (b) mesh after adaptive refinement at final simulation time. In (b), the refinement level at the rotor-stator interface is kept the same throughout the simulation, and thus we ensure that the adaptive refinement does not create mismatching cell sizes at such boundaries.}
\label{Fig-rushton}
\end{figure}

To evaluate the MEM implementation in the context of an engineering application, we simulate a two-dimensional Rushton mixer proposed by Karimian and Sewerin \cite{Karimian2025}. The dimensions indicated in\textbf{ Figure~\ref{Fig-rushton-geometry}} are detailed in \textbf{Table~\ref{Tab-rushton-params}}. This example allows us to verify the solution against existing finite-volume numerical results, as well as study the torque convergence with respect to the mesh refinement. We consider a kinematic viscosity of $\nu = 10^{-2}$~m$^2$/s, a unit density ($\rho = 1$ kg/m$^3$) and an impeller angular velocity of $\omega = 2 \pi$~rad/s; the Reynolds number is expressed as:
\begin{equation}
    Re=\dfrac{\omega d_i^2}{2 \pi \nu},
\end{equation}
\noindent and thus, for the current example, $Re=79$. The simulation is carried out for $t_{end} = 20$~s, yielding twenty revolutions to ensure that the flow is fully established at the end. The adopted initial time step is $\Delta t = 10^{-4}$~s and time step adaptation is enabled while the maximum allowed CFL is $1$. The initial mesh contains $N_e= 4,732$ biquadratic elements ($70,422$ DoFs); maintaining a fixed number of elements at the rotor-stator interface, an adaptive mesh refinement is also enabled (\textbf{Figure~\ref{Fig-rushton-mesh}}).

\begin{table}[htb!]
\centering
\caption{Dimensions of Rushton mixer tank example (Figure~\ref{Fig-rushton}) presented in Section~\ref{Sec-sub-2Drushton}.}
\begin{tabular}{clc}
    \hline
    Parameter   & Description                   & Value             \\ \hline
    $D$         & Tank diameter                 & $2$ m             \\
    $d_d$       & Disc diameter                 & $0.5$ m           \\
    $d_i$       & Impeller diameter             & $0.89$ m          \\
    $\theta_i$  & Blade opening angle           & $7.5^{\circ}$     \\
    $d_{int}$   & Mortar interface diameter     & $D/2$             \\
    $\theta_b$  & Baffle opening angle          & $\theta_i$         \\
    $l_b$       & Baffle length                 & $D/10$             \\
    \hline
\end{tabular}
\label{Tab-rushton-params}
\end{table}

\subsubsection{Unbaffled case: comparison against existing numerical results}
\label{Sec-sub-2Drushton-unbaffled}

\begin{figure}[htb!]
\centering
\begin{subfigure}[b]{0.45\textwidth}
\centering
    \includegraphics[scale=0.25]{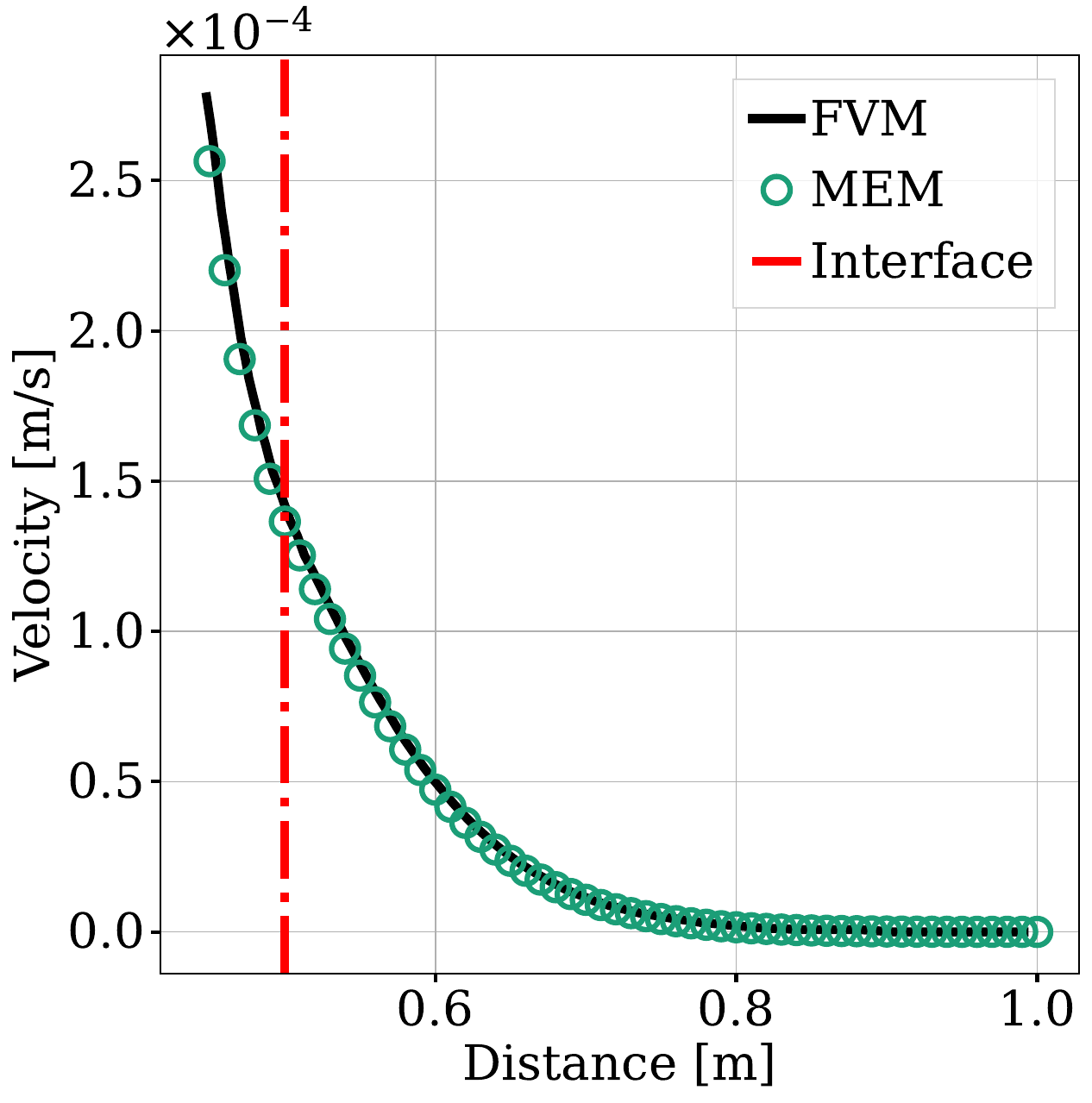}
\caption{Velocity profile at $x$-axis at $t=1$~s; unbaffled}
\label{Fig-rushton-velocity-plot}
\end{subfigure}
\begin{subfigure}[b]{0.45\textwidth}
\centering
    \includegraphics[scale=0.22]{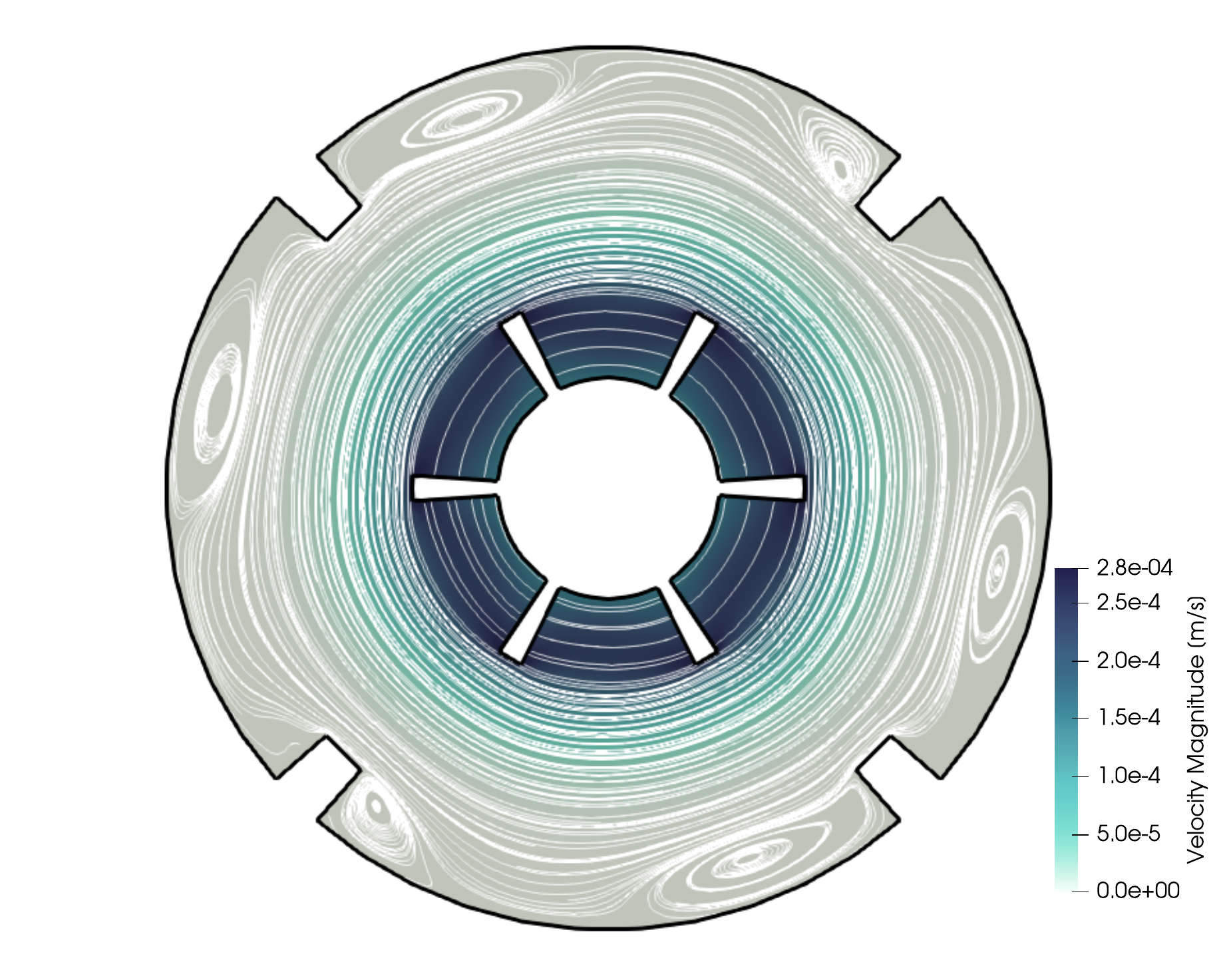}
\caption{Velocity contour at $t_{end}=20$~s; baffled}
\label{Fig-rusthon-velocity-contour}
\end{subfigure}
\caption{Two-dimensional Rushton mixer: (a) velocity profile for the unbaffled case, where the red dash-dotted line indicates the mortar interface position; (b) velocity field representation at $t_{end}$ for the baffled case.}
\end{figure}

To verify the accuracy of the mortar interface, we compare the result of an unbaffled Rushton mixer against the velocity profile reported by Karimian \textit{et al.} \cite{Karimian2025}; the mixer dimensions are as indicated in Figure~\ref{Fig-rushton-geometry}. In such study, a mesh-tying method is used in a finite-volume context, where relative rotor-stator motion is represented by a rotating non-Galilean frame of reference. \textbf{Figure~\ref{Fig-rushton-velocity-plot}} depicts the velocity profile along the $x$-axis (at $t=1$s) for both the MEM solution and the finite-volume solution obtained in \cite{Karimian2025}, where a good agreement is observed.

\subsubsection{Baffled case: torque analysis}
\label{Sec-sub-2Drushton-baffled}

The simulation of the baffled geometry represented in Figure~\ref{Fig-rushton-geometry} yields the velocity field depicted in \textbf{Figure~\ref{Fig-rusthon-velocity-contour}} at the final simulation time, when steady state is reached. We consider the baffled mixer configuration for a torque analysis study in which a uniform mesh refinement is employed. 

\begin{figure}[htb!]
\centering
    \includegraphics[width=0.9\textwidth]{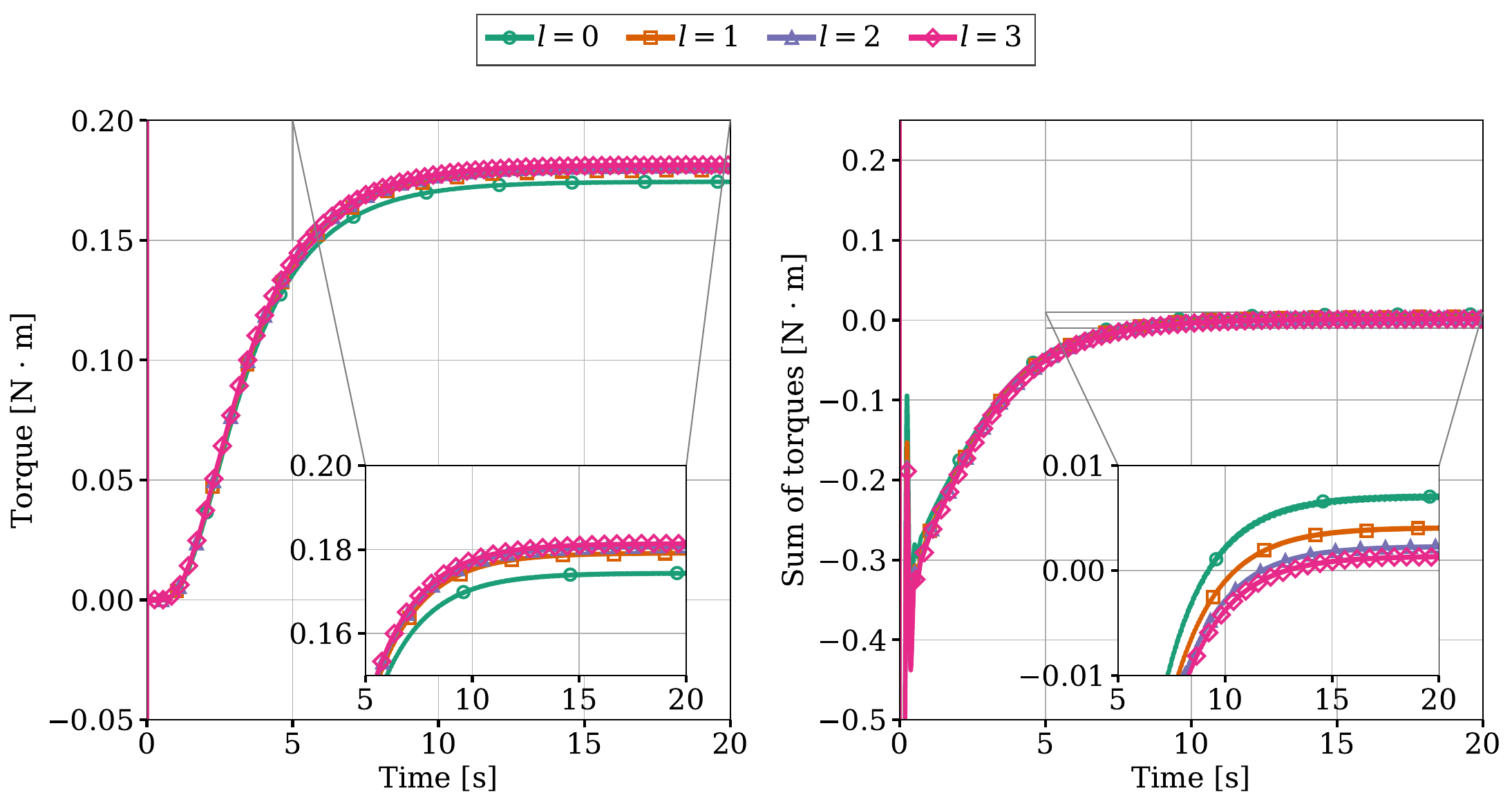}
\caption{Torque analysis for two-dimensional Rushton mixer presented in Section \ref{Sec-sub-2Drushton}: (a) torque at the impeller and (b) sum of torques at all boundaries.}
\label{Fig-rushton-torque}
\end{figure}

\textbf{Figure~\ref{Fig-rushton-torque}} depicts the torque values at the impeller and the sum of torques at all boundaries for four refinement levels; $l=0$ corresponds to $N_e=412$ elements. While mesh convergence is observed for both measurements, it is also noticeable that the sum of torques converge to zero; the latter is an expected behavior given the laminar regime of the problem. Considering the biquadratic approximation of the velocity and pressure fields, we expect a second-order torque convergence rate. We can observe the convergence trend through the change in the steady-state measured values as the mesh is refined. For instance, the variation of the computed torque at the outer wall decreases from approximately $2.7\%$ between $l=0$ and $l=1$ to $0.3$\% between $l=2$ and $l=3$.

\subsection{Three-dimensional pitched-blade turbine (PBT) mixer}
\label{Sec-sub-3Dpbt}

\begin{figure}[htb!]
\centering
    \includegraphics[scale=0.4]{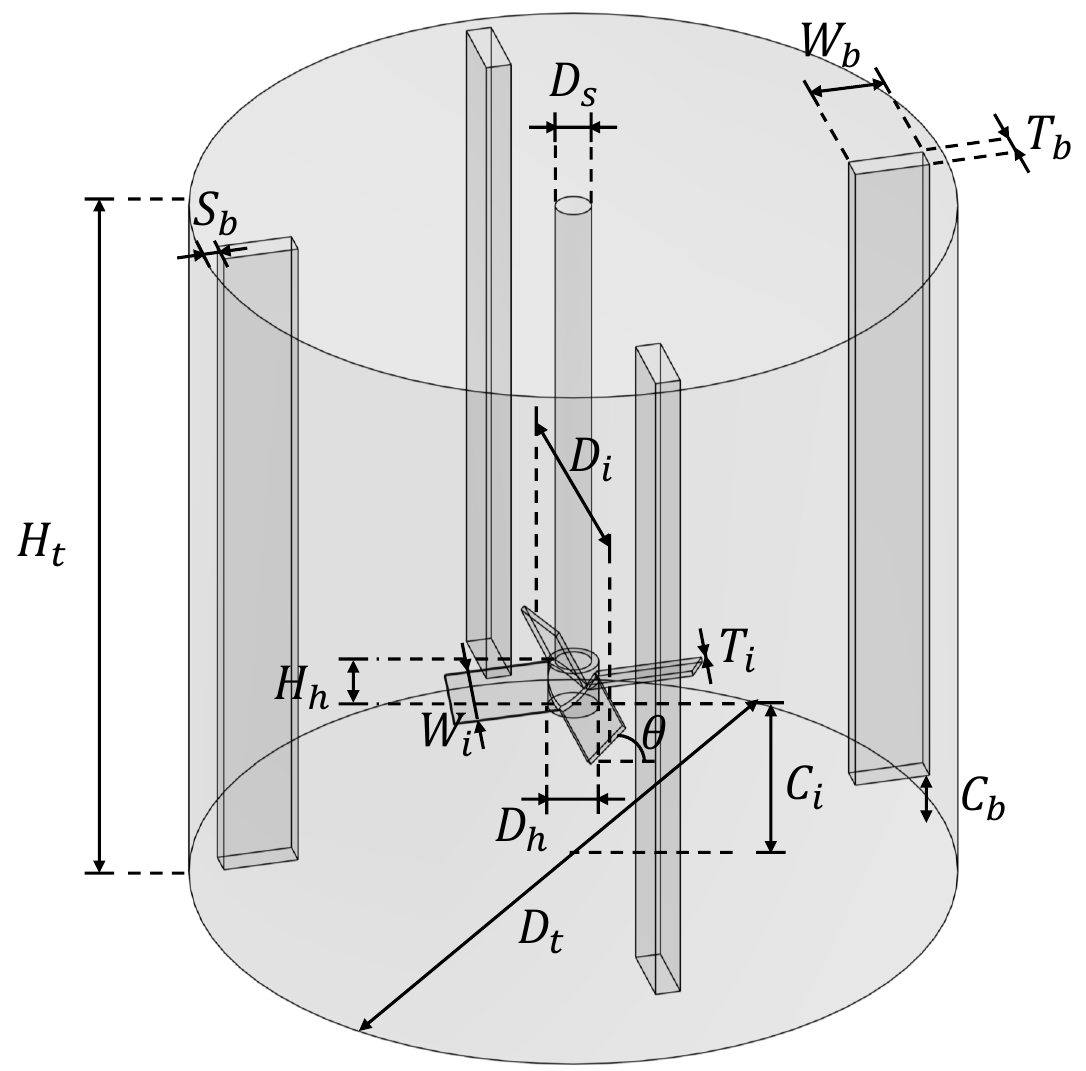}
\caption{Geometry of three-dimensional PBT mixer tank.}
\label{Fig-pbt-geometry}
\end{figure}

In this section we validate the matrix-free MEM solver with a three-dimensional pitched-blade turbine (PBT) mixer for which experimental results were reported by Blais et al. \cite{Blais2016}. The geometry is shown in \textbf{Figure~\ref{Fig-pbt-geometry}}, and the corresponding dimensions are indicated in \textbf{Table~\ref{Tab-pbt-params}}. The fluid viscosity is varied so that we can impose different Reynolds number values; the latter is computed as:
\begin{equation}
    Re = \dfrac{D_i^2 N}{\nu},
\end{equation}
\noindent where $D_i$ is the impeller diameter in [m] as reported in Table~\ref{Tab-pbt-params}, $\nu$ is the kinematic viscosity in [m$^2$/s], and $N$ is the impeller speed in [Hz]. Such rotation speed can also be represented in terms of the angular velocity $\omega = 2 \pi N$ [rad/s]. For the current problem, $N = 16$~Hz, which yields a velocity at the tip of the impeller of $u_{tip} \approx 2 \pi$~m/s. While the angular velocity is imposed at the impeller, a no-slip boundary condition is enforced on the outer cylinder, bottom wall, and baffles. A slip condition is applied to the top boundary to simulate a free fluid surface.

\begin{table}[htb!]
\centering
\caption{Dimensions of PBT mixer tank (Figure~\ref{Fig-pbt-geometry}).}
\begin{tabular}{clc}
    \hline
    Parameter   & Description                           & Value                 \\ \hline
    $D_t$       & Tank diameter                         & 0.365 m               \\
    $H_t$       & Tank height                           & $D_t$                 \\
    $C_i$       & Impeller distance to bottom           & $D_t/4$               \\
    $D_i$       & Impeller diameter                     & $D_t/3$               \\
    $W_i$       & Impeller blade width                  & $D_i/5$               \\
    $T_i$       & Impeller blade thickness              & $W_i/10$              \\
    $\theta$    & Impeller blade inclination            & $45^{\circ}$          \\
    $D_s$       & Shaft diameter                        & $W_i \cos(\theta)$    \\
    $D_h$       & Hub diameter                          & 1.4 $D_s$             \\
    $H_h$       & Hub height                            & $W_i$                 \\
    $C_b$       & Baffle distance to bottom             & $C_i/3$               \\
    $S_b$       & Baffle distance to outer tank wall    & $19/20 \times D_t$    \\
    $W_b$       & Baffle width                          & $D_t/10$              \\
    $T_b$       & Baffle thickness                      & $W_b/3$               \\
    \hline
\end{tabular}
\label{Tab-pbt-params}
\end{table}

The experimental dataset \cite{Blais2016} provides the power number $N_p$ for $Re \in [1, 2\times 10^3]$ interval. The power number is defined as:
\begin{equation}
    N_p = \dfrac{2 \pi T}{\rho N^2 D_i^5} = \dfrac{P}{\rho N^3 D_i^5},
\end{equation}
\noindent where $T$ is the torque at the impeller in [N $\cdot$ m] and $P = 2 \pi N T$ is the power consumption in [W]. 

\begin{figure}[htb!]
    \centering
    \includegraphics[width=0.9\linewidth]{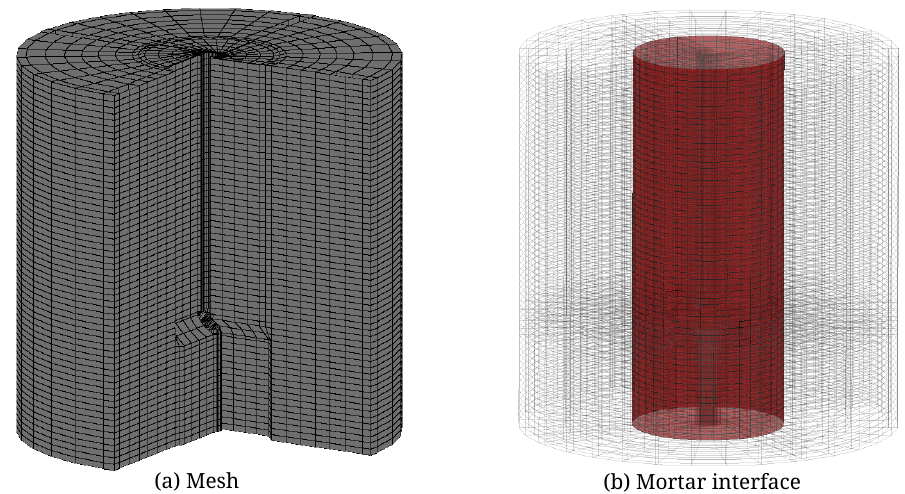}
    \caption{(a) Mesh discretization of the three-dimensional PBT mixer. The coarsest level (herein represented) contains $31,459$ cells. (b) Mortar interface (indicated in red) with diameter $D_{int} = 1.2 D_i$, where $D_i$ is the impeller diameter (Table~\ref{Tab-pbt-params}). The discretization at the interface maintains a uniform cell size in both the $x$-$y$ plane and $z$-direction.}
    \label{Fig-pbt-mesh}
\end{figure}

The problem is simulated for $t_{end} = 0.2$~s, which yields $3.2$ impeller revolutions. We consider a Newtonian fluid with unit density ($\rho = 1$ kg/m$^3$). Similarly to the two-dimensional example from Section~\ref{Sec-sub-2Drushton}, the maximum allowed CFL is 1, and the initial time step chosen is $\Delta t = 10^{-4}$~s. The mesh discretization and the mortar interface placement are represented in \textbf{Figure~\ref{Fig-pbt-mesh}}; the three refinement levels tested for different polynomial approximations are reported in \textbf{Table~\ref{Tab-pbt-mesh}}. The cell size in the radial $x$-$y$ plane and in the $z$-direction must be constant at the mortar interface, which restricts further coarsening of the initial configuration. Nonetheless, the use of a matrix-free solver yields reasonable simulation times while the high-order polynomial approximation ensures sufficient accuracy of the solution -- as discussed in the following section.

\begin{table}[htb!]
\centering
\caption{Number of cells and degrees of freedom (DoFs) for each refinement level in the three-dimensional PBT example.}
\label{Tab-pbt-mesh}
\begin{tabular}{cccc}
    \hline
     Level  & Cells & DoFs ($Q_2Q_2$)   & DoFs ($Q_3Q_3$)   \\ \hline
     0      & 32k   & 1M                & 3.6M              \\
     1      & 252k  & 8.4M              & 28M               \\
     2      & 2M    & 66M               & -                 \\ \hline
\end{tabular}
\end{table}

\subsubsection{Mesh Sensitivity}
\label{Sec-sub-pbt-convergence}

\begin{figure}[htb!]
\begin{subfigure}[b]{0.45\textwidth}
\centering
    \includegraphics[scale=0.3]{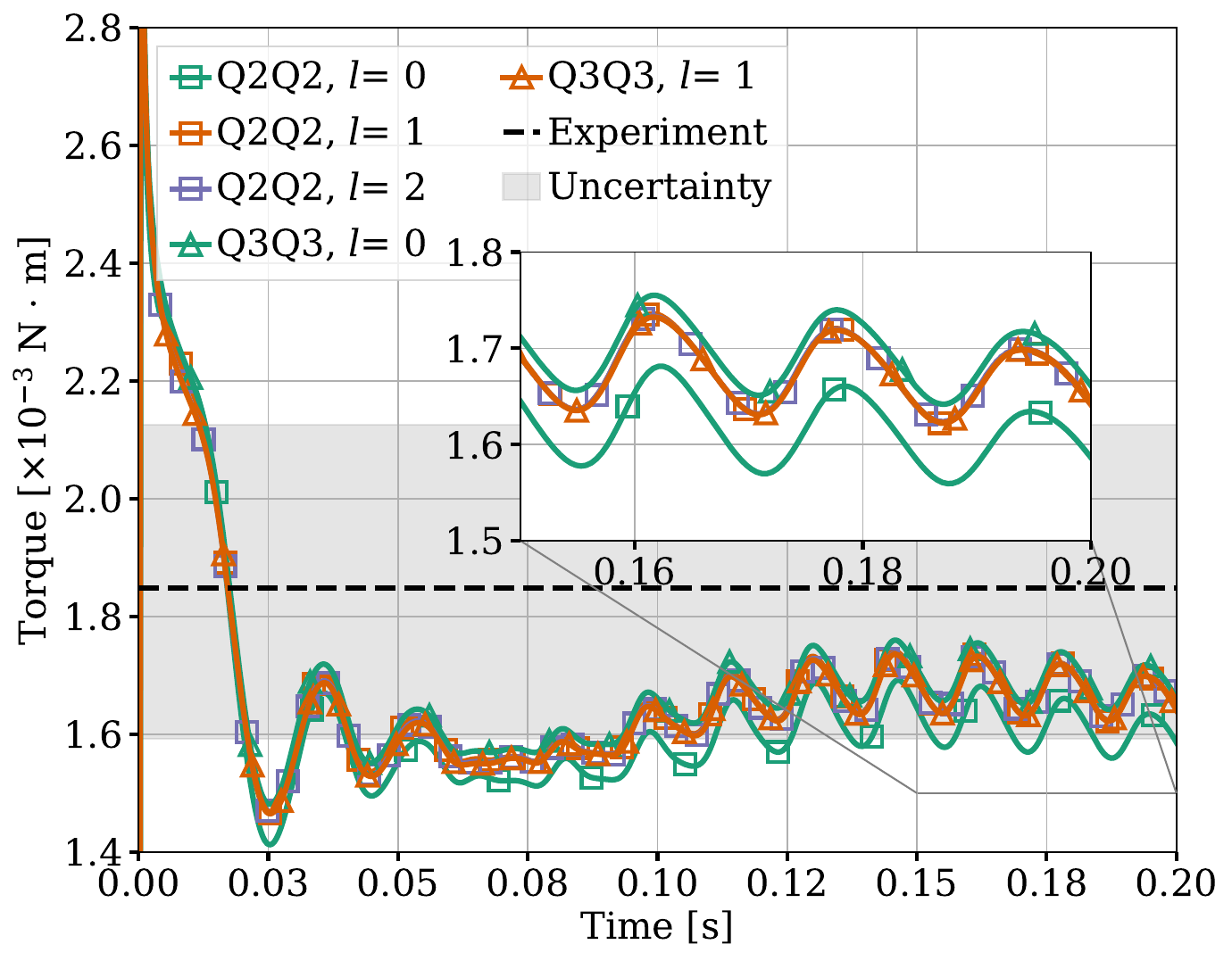}
\caption{$Re=200$}
\label{Fig-pbt-torque-Re200}
\end{subfigure}
\hfill
\begin{subfigure}[b]{0.45\textwidth}
\centering
    \includegraphics[scale=0.3]{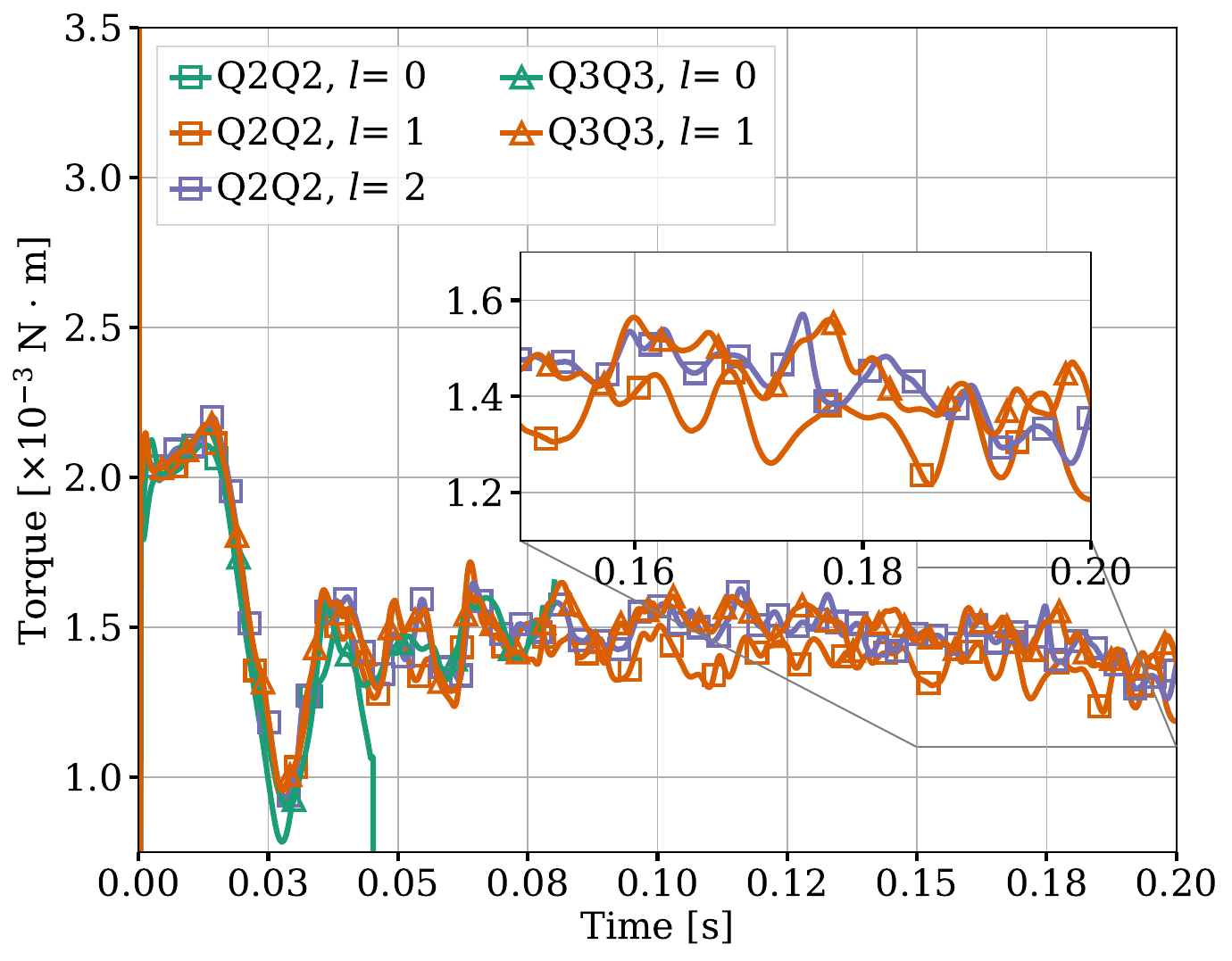}    
\caption{$Re=2000$}
\label{Fig-pbt-torque-Re2000}
\end{subfigure}
\caption{Torque analysis at the impeller for three-dimensional PBT mixer.}
\end{figure}

To verify mesh convergence, we choose a Reynolds number of $Re=200$, and hence $\nu = 1.1842 \times 10^{-3}$~m$^2$/s. For this case, we monitor the torque computed at the impeller as depicted in \textbf{Figure~\ref{Fig-pbt-torque-Re200}}. Results are shown for $Q_2Q_2$ and $Q_3Q_3$ elements; we observe that the refinement level $l=1$ of the quadratic approximation yields converged torque values. Furthermore, the variation of the computed torque at the impeller from $l=0$ to $l=1$ is approximately $3.34 \%$ for $\mathcal{P}=2$, and around $1.05 \%$ when $\mathcal{P}=3$. This indicates that the coarsest mesh level with $\mathcal{P}=2$, even with only $\approx 31$k cells, already yields fairly accurate results. Nevertheless, we choose the converged mesh configuration of $Q_2Q_2, l=1$ for the study presented in Section~\ref{Sec-sub-pbt-power}.

It is noteworthy that the torque curves, as well as the energy balance curves reported in the upcoming Section~\ref{Sec-sub-pbt-energy}, depict an oscillatory pseudo-steady state profile. Part of this oscillation is due to the angular rotation speed imposed in the problem, and is asymptotically vanishing with time and mesh refinement. The second source of oscillation is the interaction between baffles and blades within the domain, causing the solution to converge to a sinusoidal-like signal that depends on the angle between the blades and the baffles.

The mesh sensitivity study is extended for $Re=2000$, and the torque evaluated at the impeller is shown in \textbf{Figure~\ref{Fig-pbt-torque-Re2000}}. The increased turbulence at this regime leads to more erratic oscillations in the torque curves; nonetheless, results for $l=1$ and $l=2$ for $Q_2Q_2$ elements are very close, and the same is observed for $l=1$ when $\mathcal{P}=3$. Considering the same solution parameters as in $Re=200$, the results with the coarse mesh ($l=0$) for $Re=2000$ diverged at $t \approx 0.05$ s. Consequently, choosing the $Q_2Q_2, l=1$ mesh discretization still yields converged results at the upper bound of the $Re$ interval herein simulated.

\subsubsection{Energy Analysis}
\label{Sec-sub-pbt-energy}

Similarly to the energy analysis presented in \cite{Saavedra2025b} for a turbulent Taylor-Couette flow case, we evaluate the numerical dissipation inherent to the FEM model herein presented. For a domain $\Omega$, the rate of change of kinetic energy in the system normalized by the density $\rho$ is given by \cite{Bird2002}:
\begin{align}
    \int_\Omega \dfrac{\partial}{\partial t} \left( \dfrac{1}{2} u^2 \right) \: d\Omega & = - \int_{\Omega} \nabla \cdot (\dfrac{1}{2} u^2 \mathbf{u}) \: d\Omega - \int_{\Omega} \nabla \cdot (p^* \: \mathbf{u}) \: d\Omega + \int_{\Omega} p^* (\nabla \cdot \mathbf{u}) \: d\Omega \nonumber \\
    & \qquad - \int_{\Omega} \nabla (\boldsymbol{\tau}\cdot \mathbf{u}) \: d\Omega + \int_{\Omega} (\boldsymbol{\tau} : \nabla \mathbf{u}) \: d\Omega,
\label{Eq-energy-balance-no-rho}
\end{align}
\noindent where $u$ is the magnitude of $\mathbf{u}$ and $\boldsymbol{\tau} = - \nu (\nabla \mathbf{u} + \nabla \mathbf{u}^\top)$ is the deviatoric stress tensor. For a closed domain the first right-hand side (RHS) term is zero, and the third RHS term is also zero  for an incompressible flow; we can integrate by parts the remaining RHS terms. Multiplying \eqref{Eq-energy-balance-no-rho} again by the density so that we recover the physical meaning of the energy balance, the equation reduces to:
\begin{align}
    \underbrace{\int_\Omega \dfrac{\partial}{\partial t} \left( \dfrac{1}{2} \rho u^2 \right) \: d\Omega}_{\text{rate of kinetic energy}} & = - \underbrace{\int_{\partial \Omega} \mathbf{n} \cdot (\rho [\boldsymbol{\tau} + p^*]\cdot \mathbf{u}) \: d\partial \Omega}_{\text{power at the wall}} + \underbrace{\int_{\Omega} \rho (\boldsymbol{\tau} : \nabla \mathbf{u}) \: d\Omega}_{\text{viscous dissipation}}.
\label{Eq-energy-balance}
\end{align}

\begin{figure}[htb!]
\centering
    \includegraphics[scale=0.3]{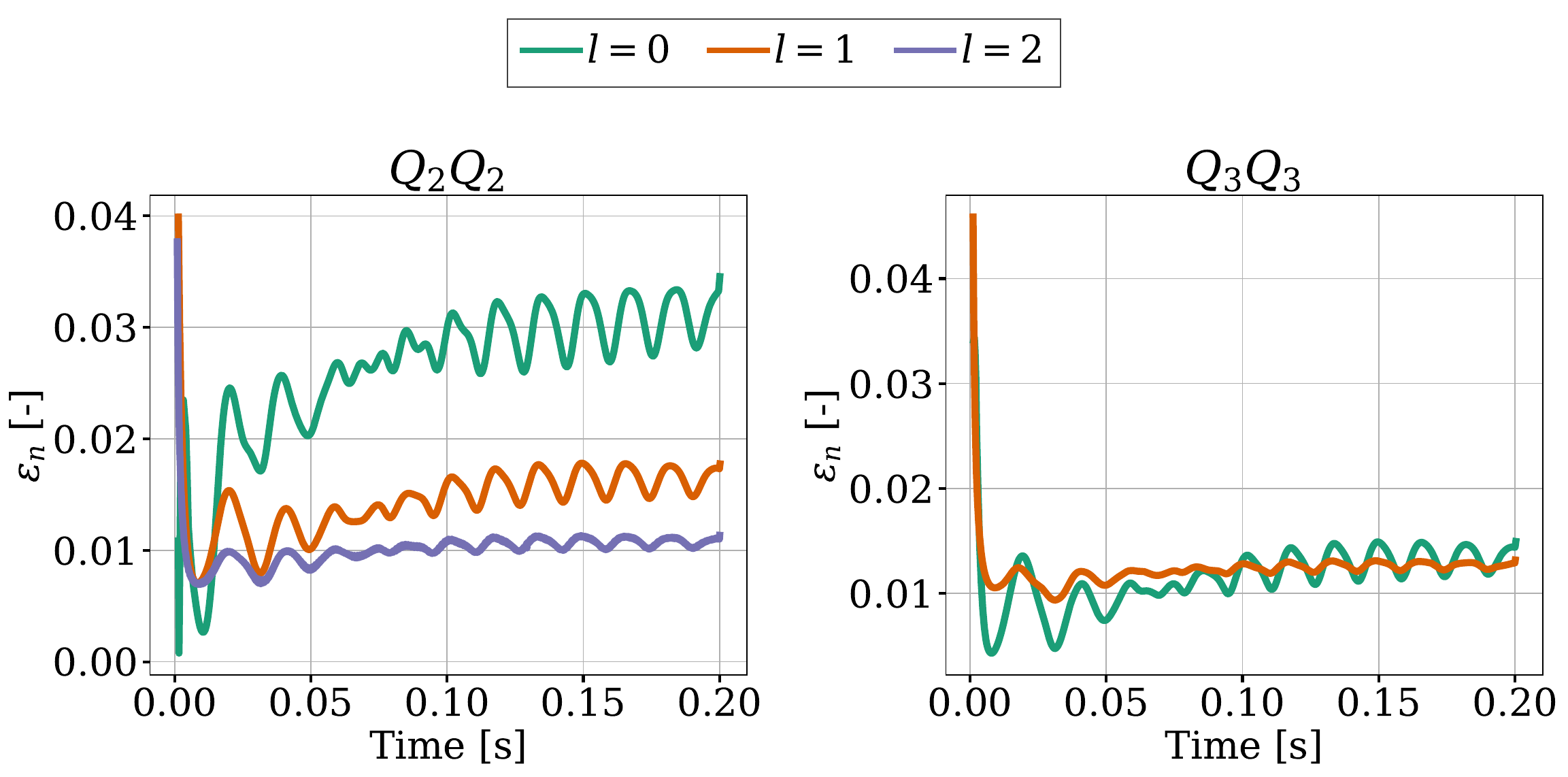}
\caption{Energy balance for three-dimensional PBT mixer at $Re=200$.}
\label{Fig-pbt-energy-Re200}    
\end{figure}

The first RHS term represents the power applied to the wall, while the second term indicates the viscous dissipation within the domain. To evaluate the magnitude of the numerical dissipation, we rewrite the energy balance in terms of a normalized positive residual $\varepsilon_n$, such that:
\begin{equation}
    - \varepsilon_n = \dfrac{1}{\overline{T}} \left( \int_{\Omega} \dfrac{\partial}{\partial t} \left(\dfrac{1}{2} u^2 \right) d\Omega + \int_{\partial \Omega} \mathbf{n} \cdot ([\boldsymbol{\tau} + p^*] \cdot \mathbf{u}) d \partial \Omega + \int_{\Omega} (-\boldsymbol{\tau} : \nabla \mathbf{u}) d\Omega \right),
\label{Eq-energy-residual}
\end{equation}
\noindent in which $\overline{T}$ is the torque power computed at $t_{end}$ for the finest mesh. \textbf{Figure~\ref{Fig-pbt-energy-Re200}} depicts $\varepsilon_n$ over time; the observed numerical dissipation is around $2\%$ for the coarsest $Q_2Q_2$ mesh, and it reaches approximately $1\%$ for the finest meshes in both $\mathcal{P}=2,3$ cases. At $t_{end} = 0.2$~s, approximately $95\%$ of the system's kinetic energy is achieved. Additional tests for a case with $Q_1Q_1$ elements showed a numerical dissipation of $\approx 10\%$, which emphasizes the advantage of the high-order approximation. 

\begin{figure}[htb!]
\centering
\begin{subfigure}[b]{0.45\textwidth}
\centering
    \includegraphics[scale=0.3]{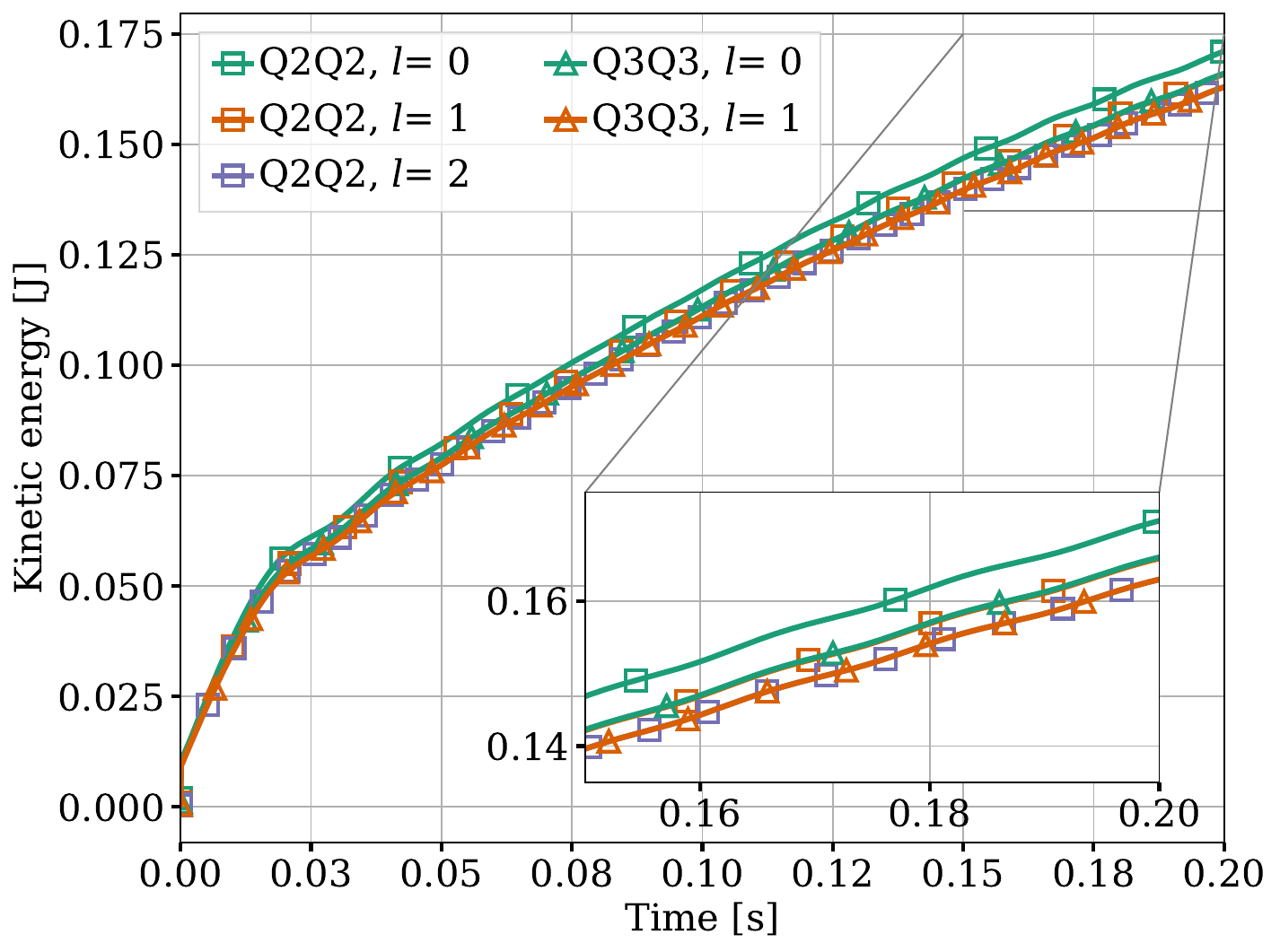}
\end{subfigure}
\hfill
\begin{subfigure}[b]{0.45\textwidth}
\centering
    \includegraphics[scale=0.3]{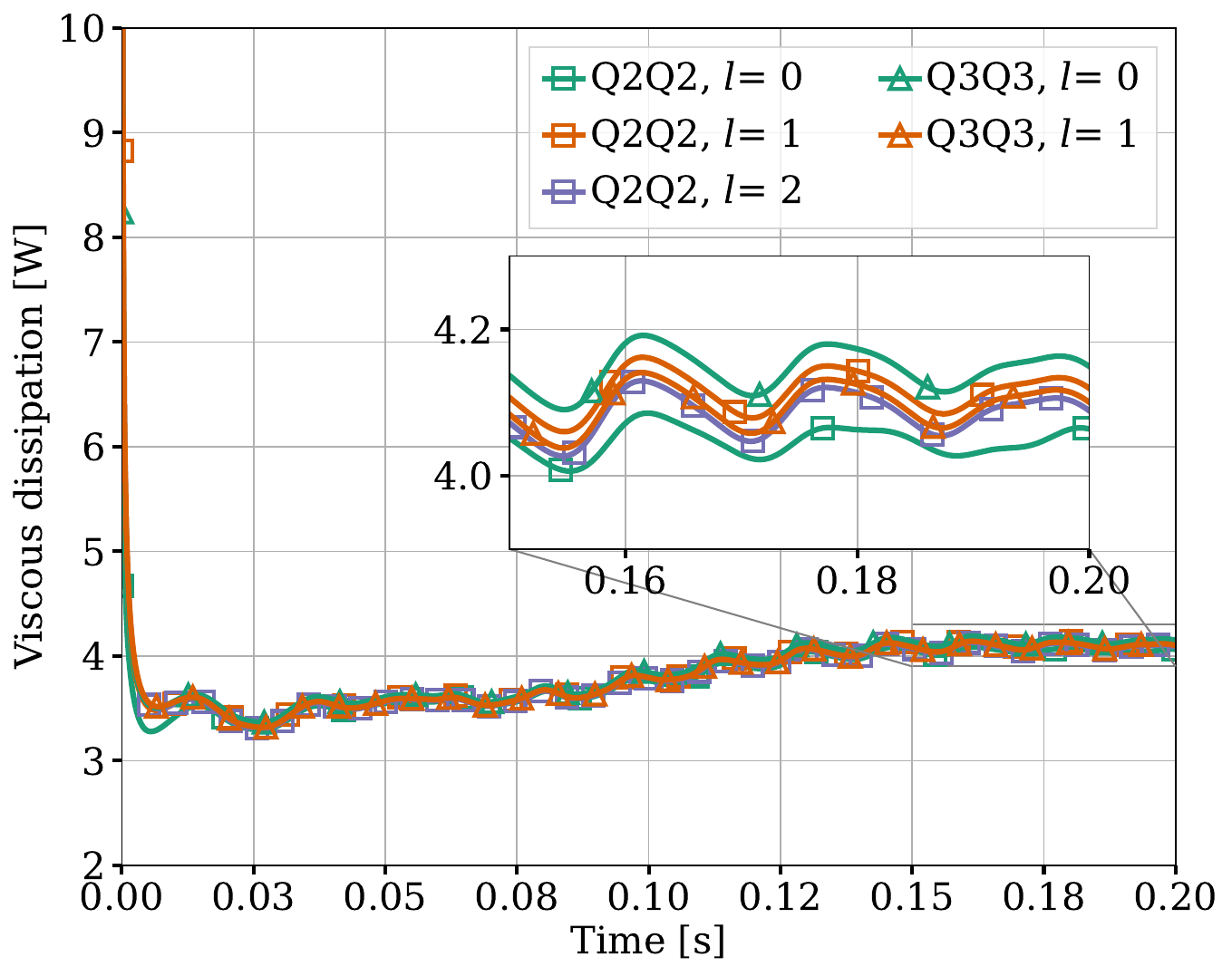}
\end{subfigure}
\caption{Kinetic energy (left) and viscous dissipation rate (right) for three-dimensional PBT mixer at $Re=200$.}
\label{Fig-pbt-ke-visc-Re200}
\end{figure}

As observed in Figure~\ref{Fig-pbt-energy-Re200}, the residual numerical dissipation levels off at around $1\%$ for both $\mathcal{P}=2$ and $\mathcal{P}=3$. To investigate the reason for this, additional tests were carried out by limiting the CFL to $0.5$ and hence reducing the maximum time step by half. This change did indeed reduce the residual numerical dissipation, meaning that part of the remaining numerical error is related to the time discretization. Furthermore, the convergence of the individual terms in Eq. \eqref{Eq-energy-residual} might affect the overall convergence pattern. To this end, \textbf{Figure~\ref{Fig-pbt-ke-visc-Re200}} depicts the kinetic energy and the viscous dissipation rate. The mesh sensitivity of the two terms differs by one order of magnitude, highlighting how the behavior of each term contributes differently to the convergence of the energy balance residual.

\begin{figure}[htb!]
\centering
    \includegraphics[scale=0.3]{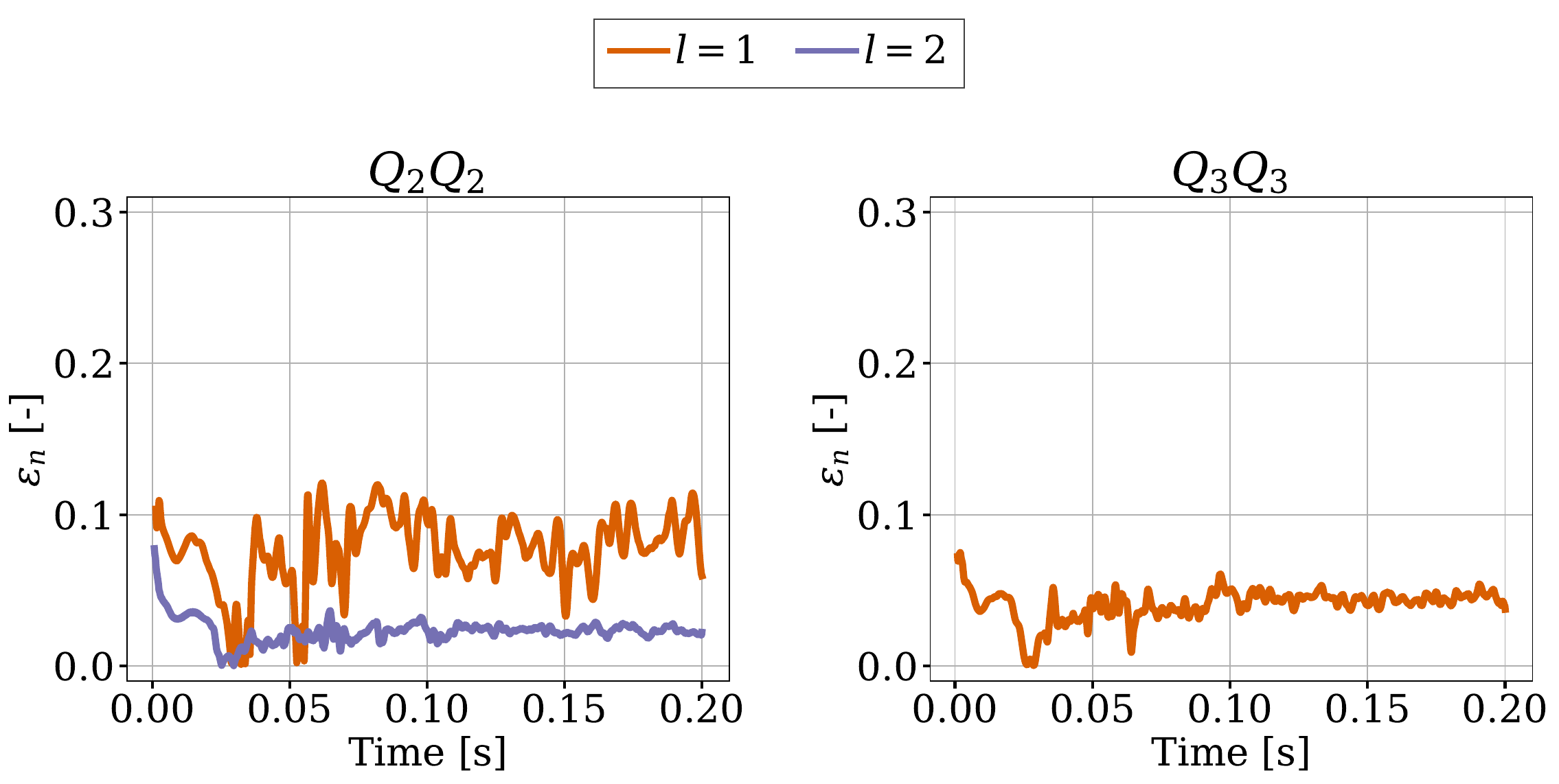}
\caption{Energy balance for three-dimensional PBT mixer at $Re=2000$.}
\label{Fig-pbt-energy-Re2000}    
\end{figure}

The energy balance is also analyzed at $Re=2000$, as depicted in \textbf{Figure~\ref{Fig-pbt-energy-Re2000}}. At this turbulence level, the coarsest mesh ($l=0$) is not able to properly represent the flow and hence the linear solver diverges quickly. Nonetheless, results for the other two mesh levels yield satisfactory numerical dissipation results: around 10\% for $l=1$ and 3\% for $l=2$ with $Q_2Q_2$ elements, and around 5\% for $l=1$ and $Q_3Q_3$ elements. Moreover, the maximum CFL allowed in the simulation is 1, and tightening this restriction would reduce the time discretization error and hence the numerical dissipation (as noted for $Re=200$). It is noteworthy that, while the mesh with refinement $l=1$ at $Re=2000$ is converged with respect to the torque metric, the energy residual remains more mesh sensitive.

\subsubsection{Power Number Curve}
\label{Sec-sub-pbt-power}

Based on the mesh convergence study and the energy analysis, in this section we obtain the power number curve for the three-dimensional PBT mixer considering a mesh with $Q_2Q_2$ elements and refinement level $l=1$ (see Table~\ref{Tab-pbt-mesh}). Twenty-four sample points were chosen in the Reynolds number interval $[1, 2\times 10^3]$; the simulations were carried out in the Trillium cluster from the Digital Research Alliance of Canada \cite{Trillium}. The simulation time decreased with the Reynolds number; for $Re \geq 100$, the average total wallclock time is 5h, while for $Re < 100$ the average running time is 8h.

The higher computational time required for low $Re$ could be associated with the laminar nature of the problem. The impeller rotation is abruptly imposed at the first time step; when this boundary condition is associated with the low $Re$ regime, which reaches a pseudo-steady state faster than at high $Re$, the linear solver requires more iterations to converge. 

\begin{figure}[htb!]
    \centering
    \includegraphics[scale=0.5]{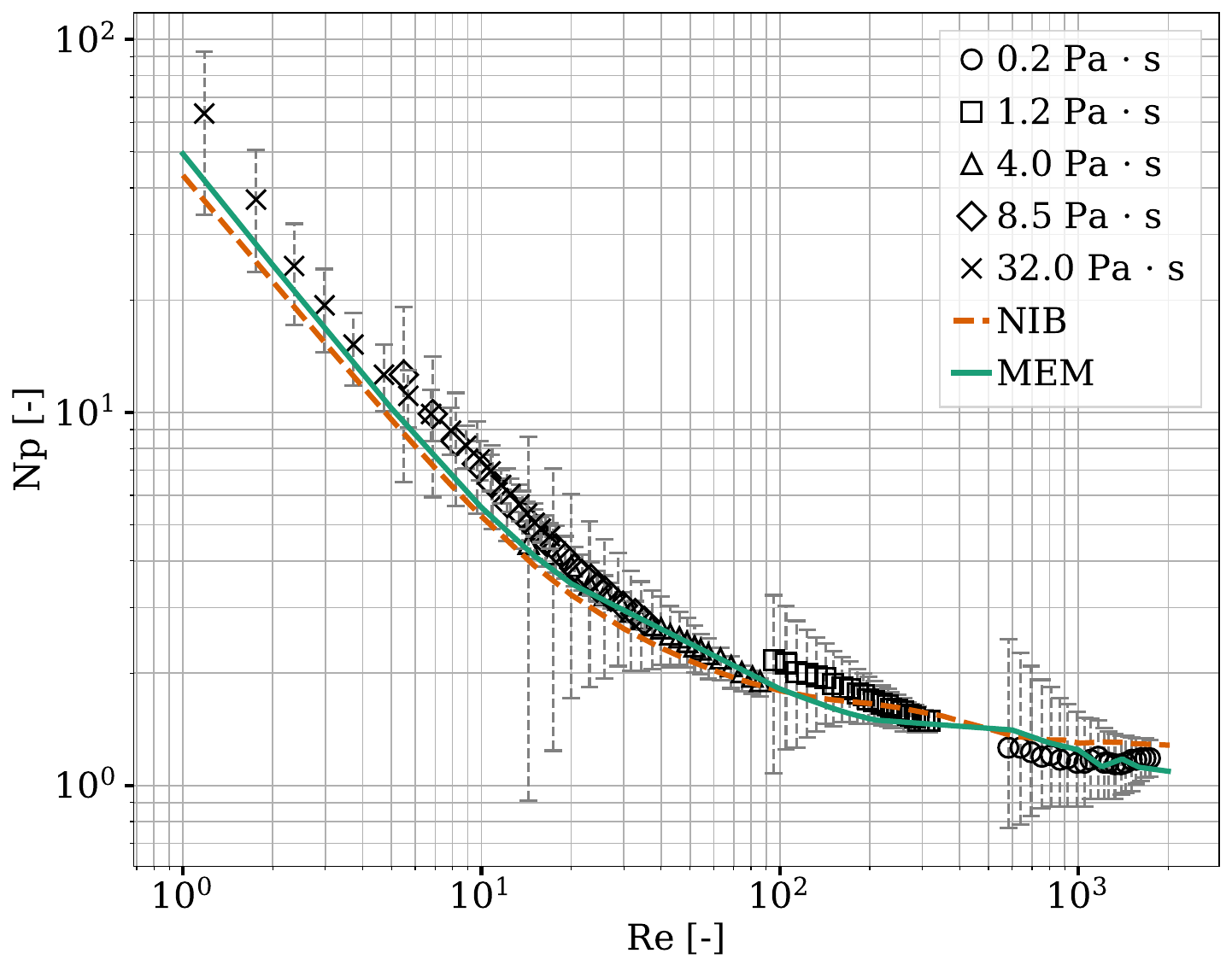}
    \caption{Power curve for three-dimensional PBT mixer; comparison of the MEM model with experimental results \cite{Blais2016} (represented by the various markers for each viscosity value) and with a Nitsche Immersed Boundary (NIB) method \cite{Joachim2023}.}
    \label{Fig-pbt-power-curve}
\end{figure}

\textbf{Figure~\ref{Fig-pbt-power-curve}} depicts the power number as a function of the Reynolds number for the experimental \cite{Blais2016} and numerical (MEM) results. A good agreement is observed, especially at higher $Re$. Previous authors reported numerical results of the same problem configuration using a Nitsche Immersed Boundary (NIB) method \cite{Joachim2023}, which are also included in Figure~\ref{Fig-pbt-power-curve}. The NIB and MEM curves are in very close agreement; the latter shows a slightly better estimation of the power number $N_p$ at $1 < Re < 100$ and at $Re > 500$ approximately. In terms of the setup and corresponding wallclock time for both methods, the NIB results reported in \cite{Joachim2023} considered a matrix-based solver with a $Q_1Q_1$ approximation. For a mesh with two million fluid cells and 120 thousand solid cells (i.e., impeller), corresponding to approximately $1.4$M DoFs in total, the study was performed in three nodes of the Narval cluster (with 64 cores each). The comparison with the MEM model parameters herein adopted is summarized in Table~\ref{Tab-pbt-compare-nib}. Even though the second-order approximation in the matrix-free solver increased significantly the number of degrees of freedom, the computational time was similar or even reduced in the MEM cases with $Re > 15$. Furthermore, the NIB simulation did not present an energy balance analysis, and hence quantification of the energy dissipation within the system was not verified.

\begin{table}[htb!]
\centering
\caption{Setup configuration and corresponding computational time for the three-dimensional PBT mixer example considering the Mortar Element Method (MEM) and the Nitsche Immersed Boundary (NIB) method results in \cite{Blais2016}. The superscripts in the NIB case indicate the number of cells in the fluid $(f)$ and solid $(s)$ domains.}
\label{Tab-pbt-compare-nib}
\begin{tabular}{ccccccc}
\hline 
Method  & Solver        & Degree    & Cells & DoFs & Cores & Time [h] \\ \hline 
MEM     & matrix-free   & $Q_2Q_2$  & $252$k & $8$M & $1 \times 192$ & 4.5 to 12\\
NIB     & matrix-based  & $Q_1Q_1$  & $2$M$^{(f)}$+$120$k$^{(s)}$ & $1.4$M  & $3 \times 64$ & 7.5 to 12.5\\ \hline
\end{tabular}
\end{table}

\section{Conclusions and further research}
\label{Sec-Conclusions}

A mortar element method within a matrix-free high-order framework was herein described and employed to simulate rotating mixing devices. The fluid domain was discretized using a CG formulation, while the mortar interface was represented with a DG framework. Continuity at such interface was imposed through a weak formulation of boundary integrals and employing a Symmetric Interior Penalty Galerkin method. An Arbitrary Lagrangian-Eulerian approach was used to account for the relative motion of the rotor and stator domains.

Two-dimensional examples were used to verify the convergence rate of the MEM matrix-free model. Both steady and transient cases yielded optimal rates for velocity and pressure fields, recovering the same order as in a standard (non-mortar) case. Regardless of the rotation angle, the geometric misalignment generated at the interface does not lead to loss of convergence properties. A three-dimensional steady case was used to evaluate the scalability of the model. When the problem is sufficiently large for the number of computing cores used, a good strong scalability is achieved. Moreover, improving the load balancing of the mortar cells did not decrease computing time significantly.

A two-dimensional Rushton mixer case validated the matrix-free MEM model against existing numerical results, and the presented torque analysis showed the expected convergence. Furthermore, a three-dimensional PBT mixer example validated the model against experimental results, where the MEM power number curve showed great agreement over the $Re$ experimental range. An energy analysis was also performed, in which the numerical dissipation introduced by the model was quantified. The converged meshes yielded $\sim 1 \%$ numerical dissipation at $Re=200$ and $10\%$ at $Re=2000$, and thus the implicit Large Eddy Simulation (LES) model herein employed is able to very accurately solve transitional/early turbulent flows. 

Future work encompasses the simulation of other devices with more complex geometries and/or flow patterns to showcase the suitability of the developed  matrix-free MEM model to a wide range of engineering applications. A current limitation of the model is the restriction of equal-sized cells at the interface. It could be advantageous, however, to allow an unstructured mesh as well as a varying interface radius so that meshing of more complex geometries is simplified; this will be addressed in future model extensions. Upcoming studies also comprise the extension of the model to couple other physics such as particle transport within a fluid.

\section*{Acknowledgments}

The authors acknowledge collaboration with the \texttt{deal.II} community, and the technical support and computing time provided by the Digital Research Alliance of Canada and Calcul Québec. The authors also acknowledge the financial support of Médicament Québec and the Quebec Consortium for Drug Discovery (CQDM) through the ARENA program, managed by the CQDM. Bruno Blais acknowledges financial support from the Natural Sciences and Engineering Research Council of Canada (NSERC) through the RGPIN-2020-04510 Discovery Grant and the funding from the Multiphysics Multiphase Intensification Automatization Workbench (MMIAOW) Canadian Research Chair Level 2 in computer-assisted design and scale-up of alternative energy vectors for sustainable chemical processes (CRC-2022-00340). 

\clearpage
\bibliographystyle{elsarticle-num} 
\bibliography{references/bibliography}

\end{document}